\documentclass[prx,amsmath,amssymb,amsbsy,twocolumn,showpacs]{revtex4-1}
\usepackage{amsmath,amssymb}
\usepackage{amsfonts}
\usepackage{wasysym}
\usepackage{graphics}
\usepackage[pdftex]{graphicx}
\usepackage{subfigure}
\usepackage{color}
\usepackage{multirow}
\usepackage{natbib}
\usepackage[breaklinks,colorlinks = true,linkcolor =blue,urlcolor=blue,citecolor=blue]{hyperref}
\usepackage{times}
\usepackage[abs]{overpic}
\usepackage{latexsym}
\usepackage{epstopdf}
\usepackage[miktex]{gnuplottex}
\usepackage[latin1]{inputenc}

\newcommand{\be}{\begin{equation}}
\newcommand{\ee}{\end{equation}}
\newcommand{\bw}{\begin{widetext}}
\newcommand{\ew}{\end{widetext}}
\newcommand{\bea}{\begin{eqnarray}}
\newcommand{\eea}{\end{eqnarray}}
\newcommand{\ba}{\begin{array}}
\newcommand{\ea}{\end{array}}

\begin{document}

\title{Bilayer mapping of the paired quantum Hall state: Instability toward anisotropic pairing}

\author{Jae-Seung Jeong}
\author{Kwon Park}
\affiliation{School of Physics, Korea Institute for Advanced Study, Seoul 130-722, Korea}

\date{\today}
\begin{abstract}
One of the most dominant candidates for the paired quantum Hall (QH) state at filling factor $\nu=5/2$ is the Moore-Read (MR) Pfaffian state. 
A salient problem, however, is that it does not occur exactly at the Coulomb interaction, but rather at a modified interaction, which favors particle-hole symmetry breaking.
In an effort to find a better state, in this work, we investigate the possible connection between the paired QH state and the antisymmetrized bilayer ground state, which is inspired by the intriguing identity that the MR Pfaffian state is entirely equivalent to the antisymmetrized projection of the bilayer QH state called the Halperin (331) state, which is valid at interlayer distance, $d$, roughly equal to the magnetic length, $l_{\rm B}$.
Specifically, by using exact diagonalization in the torus geometry, we show that the exact $5/2$ state at a given Haldane pseudopotential variation is intimately connected with the antisymmetrized bilayer ground state at a corresponding $d/l_{\rm B}$ via one-to-one mapping, which we call the bilayer mapping. 
One of the most important discoveries in this work is that the paired QH state occurring at the Coulomb interaction is mapped onto the antisymmetrized bilayer ground state at $d \gg l_{\rm B}$, which is equivalent to the antisymmetrized product state of two composite fermion seas at quarter filling, not the MR Pfaffian state.
While maintaining high overlap with the paired QH state, the antisymmetrized bilayer ground state at $d \gg l_{\rm B}$ exhibits an abrupt change under the influence of small anisotropy.
This suggests that the paired QH state occurring at the Coulomb interaction might be susceptible to anisotropic instability, opening up the possibility of anisotropic $p_x$ or $p_y$-wave pairing instead of $p_x \pm i p_y$-wave pairing in the MR Pfaffian/anti-Pfaffian state. 
\end{abstract}

\maketitle

\section{Introduction}
\label{sec:Introduction}

The fractional quantum Hall effect (FQHE) observed in the half-filled second Landau level (SLL)~\cite{PhysRevLett.59.1776,PhysRevLett.88.076801,PhysRevLett.93.176809} is one of the most fascinating phenomena in condensed matter physics, which has been attracting intense attention since its first discovery more than a quarter century ago.  
The reason for such intense attention is multifaceted.

First, occurring at filling factor $\nu=5/2$, it is the only FQHE so far observed at even-denominator fractions.
The very fact that all observed fractions for the FQHE, roughly 75 up to date~\footnote{J. K. Jain, private communication}, have odd denominators with a single exception of $\nu=5/2$ reveals that something peculiar happens in the half-filled SLL without recourse to any detailed microscopic theories.

Second, all of the above odd-denominator FQHE states are well described by a remarkably successful microscopic theory called the composite fermion (CF) theory~\cite{PhysRevLett.63.199}, which predicts the existence of an emergent quasiparticle called CF and explains the FQHE as the integer quantum Hall effect (IQHE) of composite fermions. 
Besides explaining the FQHE, the CF theory gives rise to a striking prediction that composite fermions form a Fermi sea state in the half-filled Landau levels under the assumption that composite fermions are weakly interacting~\cite{PhysRevB.46.9889,PhysRevB.47.7312}.  
This prediction was proved to be indeed the case in the half-filled lowest Landau level (LLL)~\cite{PhysRevLett.71.3850,PhysRevLett.72.2065,PhysRevLett.71.3846}. 
Unfortunately, in the half-filled SLL, this prediction is in direct contradiction with the observation of the FQHE, which requires a gap in the excitation spectrum. 
One of the natural possibilities to open a gap in the Fermi surface is through pairing between constituent fermions, in the current case, composite fermions.  
While possible, pairing between composite fermions, which themselves are composite quasiparticles containing an electron and an even number of vortices as a bound state, seems too complicated to occur.   
The possible emergence of pairing is particularly surprising from the viewpoint that the fundamental interaction between electrons is entirely repulsive and yet there is an attraction between quasiparticles. 
This is strongly reminiscent of a similar viewpoint in high-temperature superconductivity that pairing can arise from the entirely repulsive interaction in the Hubbard model.
Fortunately, in contrast to high-temperature superconductivity, where the above viewpoint still confronts various criticisms, there is a strong consensus in the FQHE community that pairing can really arise from the entirely repulsive Coulomb interaction alone.
Indeed, it has been shown that the effective interaction between composite fermions on the surface of the CF Fermi sea can become attractive in the SLL, while not in the LLL~\cite{CooperInstability}.

Third, one of the most dominant candidates describing the paired quantum Hall (QH) state is the Moore-Read (MR) Pfaffian state~\cite{moore1991nonabelions,PhysRevLett.66.3205,greiter1992paired}. 
If the MR Pfaffian state were the true ground state, the low-energy excitations of the $\nu=5/2$ state would satisfy non-Abelian braiding statistics. 
This opens up an exciting possibility of using the $\nu=5/2$ state as the solid-state quantum computer architecture that is topologically protected from various defects and errors~\cite{kitaev2003fault,PhysRevLett.94.166802,RevModPhys.80.1083}. 
Fundamentally, non-Abelian statistics originates from the existence of a zero-energy mode inside the vortex excitation, which is known to be nothing but the Majorana fermion~\cite{moore1991nonabelions,PhysRevB.61.10267,stern2010non}.
The existence of a zero-energy mode is, in turn, crucially dependent on the fact that the pairing symmetry of the MR Pfaffian state is $p_x + i p_y$, which breaks the chiral symmetry.   
Interestingly, breaking of the chiral symmetry is intimately tied with that of the particle-hole (PH) symmetry since the chiral conjugate of the MR Pfaffian state is also its PH conjugate called the anti-Pfaffian state~\cite{PhysRevLett.99.236806,PhysRevLett.99.236807}.
In this context, breaking of the PH symmetry holds an important key to the question if non-Abelian statistics can really arise in the half-filled SLL.

Fourth, the MR Pfaffian state does not respect the PH symmetry since it is the exact ground state of a model three-body interaction~\cite{moore1991nonabelions}.
While the Coulomb (or any two-body, for that matter) interaction respects the PH symmetry, the ground state can, in principle, break the PH symmetry spontaneously. 
Of course, the PH symmetry can be also broken externally in the presence of explicit symmetry breaking process.
Arguably, the most important PH-symmetry breaking process is Landau level mixing, whose precise effects have been a topic of much interest throughout recent years~\cite{PhysRevLett.106.116801,PhysRevLett.109.266806,PhysRevB.87.155426,PhysRevB.87.245129,PhysRevB.87.245425}.  
In fact, it is believed that Landau level mixing can induce an asymmetry in energy competition between the MR Pfaffian and anti-Pfaffian states, which are otherwise equal PH-conjugate partners. 
A problem is that such an asymmetry in energy competition is quite delicate. 
According to some of recent theories~\cite{PhysRevLett.106.116801,PhysRevLett.109.266806,2014arXiv1410.3861Z}, the anti-Pfaffian state might be preferred to the MR Pfaffian state. 
On the contrary, another recent theory argues that the MR Pfaffian might be favored in some range of the Landau level mixing parameter~\cite{2014arXiv1411.1068P}.

To summarize, our current understanding of the $\nu=5/2$ state is composed of two separate ingredients: 
(i) a gap opens up in the CF Fermi surface due to pairing between composite fermions, and
(ii) the resulting paired ground state is well represented by the MR Pfaffian or anti-Pfaffian state, which requires the PH-symmetry breaking.
Let us examine below how firmly each ingredient is validated.

For the first ingredient, we are interested in evidence, whose validity is free from any bias influenced by a priori assumptions about the ground state.  
Proving the existence of a gap without such bias can be achieved via numerically exact methods such as exact diagonalization (ED)~\cite{PhysRevLett.80.1505, PhysRevB.66.075408} 
and the density matrix renormalization group (DMRG) method~\cite{PhysRevLett.100.166803,2014arXiv1410.3861Z}. 
With help of recent developments in computer hardware and numerical technique, these methods can now access such large-size systems that it is possible to perform a reliable extrapolation of the Coulomb gap to the thermodynamic limit, which was estimated to be about 0.03 in units of $e^2/\epsilon l_{\rm B}$.
As a consequence, there is little doubt in that the exact $5/2$ state occurring at the Coulomb (or some slightly modified) interaction is gapped.

The existence of a gap itself, however, does not guarantee that it is due to pairing. 
To check this, by using ED in the spherical geometry, one of the current authors in collaboration with others~\cite{PhysRevB.82.201303} computed the ground state energy in the half-filled SLL as a function of particle number. 
The basic idea behind this computation is that, if pairing exists, the ground state energy should oscillates, depending on whether the particle number is even or odd. 
This is nothing but the even-odd effect, which had been used to infer the existence of pairing inside nuclei~\cite{Book05Basdevant}.  
The result obtained from ED in the half-filled SLL was that the ground state energy indeed oscillated in accordance with the existence of pairing.
To prove further that the pairing occurs between composite fermions, not between electrons, the same authors constructed and computed the superconducting order parameter for CF pairing in various finite-size systems. 
Again, the ED results showed that the superconducting order parameter for CF pairing was shown to converge into a well-defined finite value as the system sizes increases, while that for electron pairing did not.  
This provides strong evidence that pairing really occurs between composite fermions.

Now, let us switch gears to the second ingredient.
Similar to the first ingredient, we are interested in exact evidence.
Such a piece of evidence can be found in the straight overlap between the MR Pfaffian state and the exact ground state in the half-filled SLL, which has been computed in both spherical~\cite{PhysRevLett.88.216804} and torus geometries~\cite{PhysRevLett.84.4685,PhysRevB.80.241311}.
Unfortunately, the square of overlap between the two states turns out to be subpar in comparison with that between the CF state and the exact state at odd-denominator fractional filling factors, which is typically more than $99 \%$ in the finite-size systems with particle number $N \sim 10$.  
Specifically, in the spherical geometry, the square of overlap between the MR Pfaffian state and the exact ground state in the half-filled SLL is only about $70 \mbox{--} 80 \%$ in the finite-size systems with similar particle numbers~\cite{PhysRevLett.88.216804}. 
In the torus geometry, the situation gets worse in the sense that the Coulomb point is almost right on top of the phase boundary between an incompressible state, which has a sizable overlap with the MR Pfaffian state, and a stripe state, which does not~\cite{PhysRevLett.84.4685}.
In other words, the square of overlap does not even have a well-defined finite value near the Coulomb point, 
while it can be enhanced by altering the strength of the short-range part of the Coulomb interaction~\cite{PhysRevLett.84.4685,PhysRevLett.106.116801,PhysRevLett.80.1505}.
In summary, the straight overlap results obtained from ED do not seem to support the notion that the MR Pfaffian/anti-Pfaffian states represent the $\nu=5/2$ state with good accuracy, at least, at the pure Coulomb interaction.

To conduct a more specific test tailor-made to the PH-symmetry issue, one of the present authors in collaboration with others~\cite{PhysRevLett.101.156803} used  a peculiar property of the spherical geometry that the MR Pfaffian and anti-Pfaffian states occur in two separate Hilbert spaces with different particle/flux number $(N,2Q)$ combination. 
To be concrete, the MR Pfaffian and anti-Pfaffian states occur at $2Q=2N-3$ and $2N+1$, respectively, while the exact PH symmetric state occurs at the middle flux of $2Q=2N-1$.
The finite-size corrections in flux number are called the ``shifts,'' which take the values of $-3$, $+1$, and $-1$ for the MR Pfaffian, anti-Pfaffian, and the PH symmetric states, respectively.  
A consequence of these shifts is that any mixing between the MR Pfaffian/anti-Pfaffian states is artificially eliminated in the spherical geometry. 
Interestingly, this finite-size artifact can be used to test the possibility of spontaneous PH-symmetry breaking. 
That is to say, if the ground state energy shows a Mexican-hat structure as a function of particle number at a given flux number, it means that the PH symmetry can be broken spontaneously.
Unfortunately, the ED results showed that the pure Coulomb interaction did not generate the Mexican-hat structure, while an appropriately-chosen model two-body interaction could.
It is interesting to mention that the mixing between the MR Pfaffian and anti-Pfaffian states can be also suppressed in the torus geometry with hexagonal unit cell, although it can happen only for  special odd numbers of electrons~\cite{PhysRevLett.109.266806}. 
To summarize, it is unclear that the PH symmetry is spontaneously broken, at least, at the pure Coulomb or some slightly modified interaction.

As mentioned previously, the PH symmetry can be broken externally, for example, via Landau level mixing. 
In this work, we do not venture to study the effects of Landau level mixing, but focus on intrinsic properties of the ground state in the half-filled SLL.
An alternative to Landau level mixing is to modify the interaction between electrons to make the MR Pfaffian state more relevant. 
To this end, Storni and collaborators~\cite{PhysRevLett.104.076803}
had an ingenious idea of deforming the interaction between electrons in a continuous fashion from the pure Coulomb interaction to the model three-body interaction, which gives rise to the MR Pfaffian state as the exact ground state.   
If the excitation spectrum remains gapped all the way through, it may be taken as evidence for the conjecture that the MR Pfaffian state is adiabatically connected to the true ground state in the half-filled SLL.
Based on this logic, they performed ED in the spherical geometry, which showed exactly the expected result described in the above.
Unfortunately, the spherical geometry suffers from the previous mentioned finite-size artifact called the ``flux shift'' issue that any mixing between the MR Pfaffian and anti-Pfaffian states is completely suppressed. 
This means that the above result alone cannot establish the adiabatic connection between the MR Pfaffian state and the exact Coulomb ground state. 
In particular, the anti-Pfaffian state has exactly the same ``adiabatic connection'' with the exact Coulomb ground state in its own particle/flux number sector. 
The main question is which state is adiabatically connected with the true Coulomb ground state in the thermodynamic limit, where the mixing between the two states is allowed.
Again, this question is related with the attainability of spontaneous PH-symmetry breaking, which is uncertain at least at the pure Coulomb or some modified interaction according to the result mentioned in the preceding paragraph.

Gathering all pieces of evidence so far, we arrive at the following conclusion: (i) the FQHE gap at $\nu=5/2$ is induced via pairing of composite fermions, but (ii) it is unclear that the $5/2$ state occurring at the Coulomb interaction is well described by the MR Pfaffian or anti-Pfaffian state. 
In this work, we would like to improve upon the latter part of the conclusion. 
Specifically, we propose an alternative possibility that the MR Pfaffian and anti-Pfaffian states are mixed together to generate a new paired QH state with anisotropic pairing. 
This proposal is in part based on a theoretical idea that mixing the MR Pfaffian and anti-Pfaffian states, which have $p_x \pm i p_y$-wave paring, can restore the PH symmetry in exchange of the broken rotational symmetry, which results in an anisotropic paired QH state with $p_x$ or $p_y$-wave pairing.    
This idea is also motivated by the previously mentioned numerical observation~\cite{PhysRevLett.84.4685,PhysRevB.80.241311} that the Coulomb point is almost right on top of the phase boundary between the isotropic incompressible state, which is well described by the MR Pfaffian/anti-Pfaffian states, and the anisotropic compressible state, which is well described by the stripe state.   
This suggests that a certain state with both anisotropic and pairing nature could be relevant at the Coulomb interaction.

More important, our proposal is motivated by a tantalizing experimental observation by Liu {\it et al.}~\cite{PhysRevB.88.035307}; while the $\nu=5/2$ state exhibits a strong resistance anisotropy under an in-plane magnetic field, the resistances in both directions follow exactly the same Arrhenius-type temperature dependence with a single energy gap.  
The very fact that the application of an in-plane magnetic field can induce a strong resistance anisotropy in the $\nu=5/2$ state has been well known for a number of years~\cite{PhysRevLett.83.820,PhysRevLett.83.824,PhysRevLett.113.076803}. 
While the initial goal of the in-plane magnetic field experiment was to investigate the spin polarization of the $5/2$ state~\cite{PhysRevLett.61.997}, it had been realized that, in combination of finite sample thickness, the in-plane magnetic field played a much more complicated role
than just increasing the Zeeman splitting energy~\cite{PhysRevLett.110.206801,PhysRevB.87.245315}. 
In general, the in-plane magnetic field was believed to weaken the $5/2$ FQHE state by generating various anisotropic effects, which eventually induce a transition to the anisotropic compressible states observed in the third or higher Landau levels~\cite{PhysRevLett.82.394}, which are believed to be the stripe states~\cite{PhysRevLett.76.499,PhysRevB.54.1853,PhysRevB.54.5006,PhysRevLett.83.1219,PhysRevLett.85.5396}. 
An important point in the experiment by Liu {\it et al.} is that, despite such a strong resistance anisotropy, the $5/2$ FQHE state can maintain its integrity as a single incompressible state in sharp contrast to the compressible stripe state. 
Actually, the resistance anisotropy is already present even in the complete absence of an in-plane magnetic field, in which situation it is believed to be induced by an anisotropy in mobility. 
Considering that other FQHE states in the LLL do not exhibit such a sensitive resistance anisotropy depending on the mobility anisotropy, this suggests that the resistance anisotropy at $\nu=5/2$ may be an inherent property of the $5/2$ state, which can be induced by any means breaking the rotational symmetry.  
Interestingly, there is a recent experiment showing that the in-plane magnetic field can even strengthen the $\nu=5/2$ FQHE in some samples~\cite{PhysRevLett.108.196805}, corroborating the above suggestion.

In fact, a similar coexistence between the quantized Hall resistance and the resistance anisotropy had been observed at $\nu=7/3$~\cite{xia2011evidence}, which was then interpreted as a signature for the nematic QH state~\cite{musaelian1996broken,fogler2004effective,PhysRevB.82.085102}.
While Liu {\it et al.} followed the same interpretation, we think that it may well be interpreted as a signature for the anisotropic paired QH state.
In our interpretation, the single energy gap, or activation temperature determined from the Arrhenius-type temperature dependence is related with the anisotropic pairing gap in the similar fashion, in which the critical temperature of a cuprate high-temperature superconductor is related with its anisotropic $d$-wave pairing gap.

\begin{figure*}[]
\includegraphics[width=1.5\columnwidth]{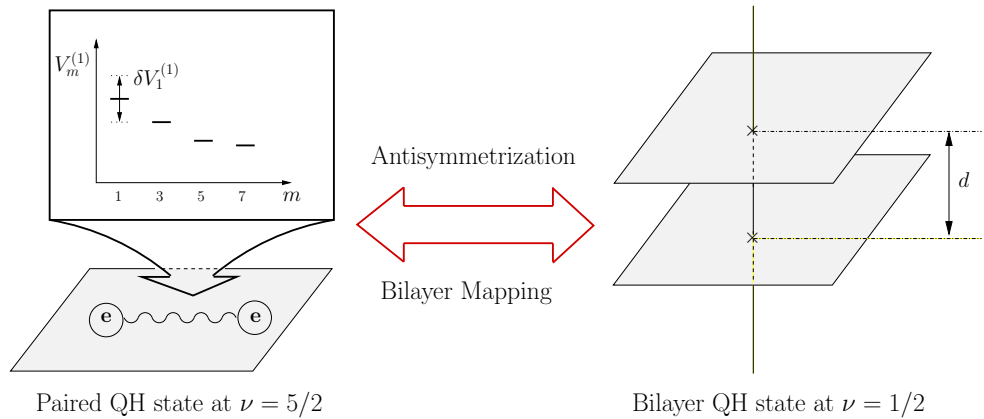}
\caption{
Schematic diagram for the bilayer mapping, via which the exact $5/2$ state at a given Haldane pseudopotential variation, $\delta V^{(1)}_1$, is mapped onto the antisymmetrized bilayer ground state at a corresponding interlayer distance, $d$.
}
\label{fig:Schematic}
\end{figure*}

Eventually, the main goal of this work boils down to constructing a trial state in the half-filled SLL, which works better than the MR Pfaffian/anti-Pfaffian states at the Coulomb interaction.
Based on the fact that the exact $5/2$ state is paired at the Coulomb interaction, such a trial state would provide a better wave function for the paired QH state.
As a next step, we then show that such a trial state is anisotropic, or sufficiently susceptible to anisotropic instability in a finite-size system.

Constructing a better trial state than the MR Pfaffian/anti-Pfaffian states is actually very difficult, which is after all one of the main reasons why the MR Pfaffian/anti-Pfaffian states have remained as the dominant candidates for the $\nu=5/2$ state for more than twenty years.   
To begin with, the MR Pfaffian/anti-Pfaffian states have the lowest energy among various known trial states, which include the CF sea states with fully polarized/unpolarized spins, the Haldane-Rezayi spin-singlet paired state~\cite{PhysRevLett.60.956}, and so on~\cite{PhysRevB.58.R10167}.
An important clue to improve upon the MR Pfaffian state may be obtained in the interpretation that the MR Pfaffian state is the BCS paired wave function of composite fermions. 
M\"{o}ller and Simon~\cite{PhysRevB.77.075319} took this interpretation seriously and proposed a trial wave function scheme for the BCS paired CF states by combining the concept of the BCS Hall states~\cite{PhysRevB.61.10267} with the explicit construction of the CF wave functions~\cite{PhysRevB.55.R4895}. 
In this scheme, the BCS paired CF wave functions can be constructed such that it is adiabatically connected to the MR Pfaffian state via careful tuning of multiple variational parameters.

We take a different approach. 
Our approach is inspired by the intriguing identity~\cite{PhysRevB.46.9586} that the MR Pfaffian state is entirely equivalent to the antisymmetrized projection of the bilayer QH state known as the Halperin (331) state, which is valid at interlayer distance, $d$, roughly equal to the magnetic length, $l_{\rm B}$~\cite{PhysRevB.39.1932}.   
Motivated by this identity, we ask if the exact $5/2$ state occurring at the Coulomb interaction can be mapped onto the antisymmetrized bilayer ground state at appropriately chosen $d/l_{\rm B}$.
If so, one can generalize this idea and think of a general mapping between the exact $5/2$ state at a given Haldane pseudopotential variation and the antisymmetrized bilayer ground state at a corresponding $d/l_{\rm B}$.
We call this mapping the bilayer mapping. 
See Fig.~\ref{fig:Schematic} for a schematic diagram for the bilayer mapping.
In essence, we ask if one can construct a trial wave function scheme based on the bilayer mapping with $d/l_{\rm B}$ serving as a variational parameter, which works better than the MR Pfaffian/anti-Pfaffian states and the BCS paired CF states mentioned in the above.

One of the most appealing properties of our variational scheme is that it is built to recover the MR Pfaffian state at $d/l_{\rm B} \simeq 1\mbox{--}2$ without any detailed tuning processes and therefore guaranteed to capture the known pairing mechanism described by the MR Pfaffian state in this parameter regime. 
Meanwhile, at $d/l_{\rm B} \ll 1$, the bilayer QH system reduces to the single-layer QH system with slightly broken SU(2) spin rotational symmetry, where the ground state is shown to be the CF sea state with mixed spin polarizations.   
It is shown in Sec.~\ref{sec:ABGS} that the antisymmetrization operator projects such a CF sea state into the fully spin-polarized sector. 
As a consequence, our scheme gives rise to a natural variational scheme interpolating between the MR Pfaffian state occurring at $d/l_{\rm B} \simeq 1\mbox{--}2$ and the fully spin-polarized CF sea state occurring at $d/l_{\rm B} \ll 1$.

Surprisingly, our scheme reveals an unexpected result that the exact $5/2$ state occurring at the Coulomb interaction is mapped onto the antisymmetrized bilayer ground state at $d/l_{\rm B} \gg 1$, not at $d/l_{\rm B} \simeq 1.5\mbox{--}2$, where the MR Pfaffian state is relevant. 
It is shown in Sec.~\ref{sec:Bilayer_mapping} that, at the Coulomb point,
the square of overlap between the exact $5/2$ state and the antisymmetrized bilayer ground state is maximal at $d/l_{\rm B} \gg 1$. 
After PH symmetrization, the square of overlap is enhanced substantially to become above $90 \%$ in the $N=12$ system, which is much larger than that ($\simeq 60 \%$) between the exact $5/2$ state and the PH-symmetry-restored MR Pfaffian state in the same finite-size system.

Microscopically, the bilayer ground state at $d/l_{\rm B} \gg 1$ is given by the product state of two CF seas at quarter filling~\cite{PhysRevB.64.085313}, which is susceptible to anisotropic instability since it is not only compressible in each pseudospin sector, but also subject to additional energy degeneracies due to the negligible interlayer interaction. 
This leads us to conjecture that the antisymmetrized bilayer ground state at $d/l_{\rm B} \gg 1$ might also be susceptible to anisotropic instability.
To check this conjecture, in Sec.~\ref{sec:Anisotropic_instability}, we investigate the behavior of the antisymmetrized bilayer ground state at $d/l_{\rm B} \gg 1$ as a function of aspect ratio of the unit cell in the torus geometry as well as physical anisotropy parameters such as the anisotropy ratio of the band mass and the dielectric tensor of the Coulomb interaction.

The rest of the paper is organized as follows.
In Sec.~\ref{sec:Hamiltonian}, we present the precise mathematical forms of the Hamiltonians for the half-filled SLL and the bilayer QH system. 
In Sec.~\ref{sec:Trial_states}, we discuss various trial states for each Hamiltonian, which are valid in particular parameter regimes.
In Sec.~\ref{sec:ABGS}, we explain how to perform the antisymmetrization operation onto the exact bilayer ground state that is obtained via ED and represented in terms of the basis states in second quantization. 
To check the validity of our antisymmetrization process, we test if the antisymmetrized bilayer ground state can reproduce the known trial states in their respective regimes. 
In particular, the antisymmetrized Halperin (331) state is shown to be precisely identical to the MR Pfaffian state, which is independently obtained as the exact ground state of the three-body interaction. 
In Sec.~\ref{sec:Bilayer_mapping}, we show that there is an one-to-one mapping between the exact $5/2$ state at a given Haldane pseudopotential variation and the antisymmetrized bilayer ground state at a corresponding $d/l_{\rm B}$, which we call the bilayer mapping.  
Through this bilayer mapping, it is shown that the exact $5/2$ state occurring at the Coulomb interaction is intimately connected with the antisymmetrized bilayer ground state at $d/l_{\rm B} \gg 1$, not the MR Pfaffian state.
In Sec.~\ref{sec:Anisotropic_instability}, we provide evidence that the antisymmetrized bilayer ground state at $d/l_{\rm B} \gg 1$ is susceptible to anisotropic instability.  
Finally, we conclude in Sec.~\ref{sec:Conclusion} by discussing how the antisymmetrization operation can in principle generate the gap in the antisymmetrized bilayer ground state at $d/l_{\rm B} \gg 1$, which is otherwise compressible.

\section{Hamiltonian}
\label{sec:Hamiltonian}

Let us begin by writing the Hamiltonian for the fully spin-polarized electrons confined in the Landau level with index $n$ ($n$LL) as follows~\cite{Book87Prange}:
\begin{align}
\label{eq:H_LLn}
H_{n{\rm LL}} = {1\over N_{\Phi}} \sum_{\bf q} V_{\bf q} e^{-{q^2/2}} L_n^2\left({q^2\over 2}\right)\sum_{i<j}e^{i{\bf q}\cdot({\bf R}_i-{\bf R}_j)},
\end{align}
where ${\bf q}=(q_x, q_y)=(2\pi s/a, 2\pi t/b)$ $\left[s, t\in {\mathbb Z}\right]$ and $q=|{\bf q}|=(q_x^2+q_y^2)^{1\over 2}$.
Here, $a$ and $b$ are the linear lengths of the rectangular unit cell along the $x$ and $y$ directions, respectively, in terms of which the aspect ratio is defined to be $r_a=a/b$. 
$V_{\bf q}$ is the Fourier transform of the interaction potential between electrons, $L_n(x)$ is the Laguerre polynomial, and ${\bf R}_i$ is the guiding center coordinates of the $i$-th electron. 
Note that $V_{{\bf q}}/(e^2/\epsilon l_{\rm B}) = 2\pi /q l_{\rm B}$ for the pure Coulomb interaction. 
From now on, the magnetic length, $l_B$, is set equal to unity throughout this paper, unless stated otherwise.
In this work, we are interested in the situation where a given Landau level with index $n$ is half filled. 
In this situation, the total flux, $N_{\Phi}$, in units of flux quantum is related with the total number of electrons, $N$, via $N_{\Phi}=2N$. 
To avoid confusion, remember that we use the convention of calling the Landau level with index $n$ the $(n+1)$-th Landau level; 
in this convention, 0LL denotes the lowest, or first Landau level (LLL), 1LL denotes the second Landau level (SLL), and so on.

The above Hamiltonian can be alternatively expressed in terms of the Haldane pseudopotential parameter, $V_m$, as follows:
\begin{align}
\label{eq:H_LLn_pseudo}
H_{n{\rm LL}} = \sum_{m=0}^{\infty}{2 V^{(n)}_m\over N_{\Phi}}\sum_{{\bf q}}e^{-{q^2/2}} L_m(q^2)\sum_{i<j}e^{i{\bf q}\cdot ({\bf R}_{i}-{\bf R}_{j})},
\end{align}
where $V^{(n)}_m$ is the Coulomb energy cost of an electron pair confined in the $n$LL with relative angular momentum, $m$. 
Mathematically, $V^{(n)}_m$ is defined as follows~\cite{Book87Prange}:
\begin{align}
\label{eq:V_m}
V^{(n)}_m={1\over 2\pi}\int d^2{\bf q} ~e^{-q^2} L_m(q^2) L_n^2\left( q^2 \over 2 \right) V_{\bf q} .
\end{align}
The Haldane pseudopotential formalism is convenient since it provides a systematic scheme for tuning the interaction strength by varying $V^{(n)}_m$ in the vicinity of the pure Coulomb values.
Often, the most important parameter is the variation in the $m=1$ component:
\begin{align}
\label{eq:delta_V_1}
\delta V^{(n)}_1 = V^{\prime(n)}_1 - V^{(n)}_1 ,
\end{align}
where $V^{\prime(n)}_1$ and $V^{(n)}_1$ are the varied and the pure Coulomb value of the Haldane pseudopotential at $m=1$, respectively.
In this work, we are mainly interested in the second Landau level, i.e., 1LL so that the most important variational parameter is $\delta V^{(1)}_1$.  
For reference, $V^{(1)}_1 / (e^2/\epsilon l_{\rm B})=0.415419$ and it is sufficient to vary $\delta V^{(1)}_1/V^{(1)}_1$ within the range from $-0.3$ to $0.3$ for the purpose of this work.

The Hamiltonian for the bilayer QH system is defined by assigning different interaction potentials for the intralayer and interlayer interaction, which is written in position space as follows:
\begin{align}
\frac{{\cal V}_{\rm intra}({\bf r})}{e^2/\epsilon l_{\rm B}} &= \frac{1}{r} ,
\nonumber \\
\frac{{\cal V}_{\rm inter}({\bf r})}{e^2/\epsilon l_{\rm B}} &= \frac{1}{\sqrt{r^2+d^2}} ,
\end{align}
where $d$ is the interlayer distance.
In the momentum space, the interaction potentials are given by
\begin{align}
\frac{V_{\rm intra}({\bf q})}{e^2/\epsilon l_{\rm B}} &= \frac{2\pi}{q} ,
\nonumber \\
\frac{V_{\rm inter}({\bf q})}{e^2/\epsilon l_{\rm B}} &= 2\pi \frac{e^{-qd}}{q} ,
\end{align}
in terms of which the Hamiltonian for the bilayer QH system, $H_{\rm BQH}$, is written in the 0LL as follows:
\begin{align}
\label{eq:H_BQH}
H_{\rm BQH} =  {1\over N_{\Phi}} \sum_{\bf q} L^2_{n=0} \left(\frac{q^2}{2}\right) e^{-{q^2/2}} \sum_{i<j}  V^{\sigma_i \sigma_j}_{\bf q}
e^{i{\bf q}\cdot({\bf R}_i-{\bf R}_j)} ,
\end{align} 
where $V^{\sigma_i \sigma_j}_{\bf q}=V_{\rm intra}({\bf q})$ if $\sigma_i$ and $\sigma_j$ belong to the same layer index, or pseudospin,  
and $V^{\sigma_i \sigma_j}_{\bf q}=V_{\rm inter}({\bf q})$ otherwise.
We are interested in solving $H_{\rm BQH}$ in the situation where the total filling factor is half; $\nu_{\rm tot}=\nu_\uparrow+\nu_\downarrow =1/2$ with both $\nu_\uparrow$ and $\nu_\downarrow$ being $1/4$.
We ignore the electron tunneling between different layers, which is present in actual bilayer QH systems.

It is important to note that, while it is possible to conduct a comprehensive study for various Landau levels, in this work, we only focus on the 0LL as far as the bilayer QH system is concerned, which means that $L^2_{n=0}(q^2/2)=1$ in Eq.~\eqref{eq:H_BQH}.
Eventually, this means that we are interested in the possible connection between the exact ground state in the half-filled SLL, which is known to be paired, and the antisymmetrized bilayer ground state in the half-filled LLL.
Put in another way, we would like to investigate if one can construct a good trial wave function scheme, based on the connection between the paired quantum Hall state and the antisymmetrized bilayer ground state.  
If so, the interlayer distance can serve as a variational parameter for the trial wave function.

In this work, we obtain both the exact ground state in the half-filled SLL and that of the bilayer QH system in the half-filled LLL by using ED in the torus geometry. 
To this end, it is necessary to rewrite the $n$LL Hamiltonian in Eq.~\eqref{eq:H_LLn} in terms of the torus basis states in second quantization: 
\begin{align}  
H_{n{\rm LL}} = \sum_{j_1,j_2,j_3,j_4}  M^{(n)}_{j_1j_2j_3j_4}c_{j_1}^{\dagger}c_{j_2}^{\dagger}c_{j_3}c_{j_4} ,
\end{align}
where $c_j$ and $c_j^{\dagger}$ are the annihilation and creation operators acting on the $j$-th state, respectively. 
The matrix element, $M^{(n)}_{j_1 j_2 j_3 j_4}$, is given as follows~\cite{PhysRevLett.50.1219}:
\begin{align}
\label{eq:Matrix_element}
M^{(n)}_{j_1 j_2 j_3 j_4}
&=\delta'_{j_1\!-j_4,t}\delta'_{j_1+j_2,j_3+j_4}
\nonumber \\
&\times \sideset{}{'}\sum_{\bf q} {V_{\bf q}\over 2N_{\Phi}} e^{iq_x(X_{j_1}-X_{j_3})}
 e^{-{q^2\over 2}} L_n^2\left({q^2\over 2}\right),
\end{align}
where $(q_x,q_y)=(2\pi s/a, 2\pi t/b)$ $\left[s, t\in {\mathbb Z}\right]$.
Alternatively, the matrix element can be written in terms of the Haldane pseudopotential as follows:
\begin{align}
\label{eq:Matrix_element_via_V_m}
M^{(n)}_{j_1 j_2 j_3 j_4}
&=\delta'_{j_1\!-j_4,t}\delta'_{j_1+j_2,j_3+j_4}
\nonumber \\
&\times \sum_{m=0}^\infty \frac{V^{(n)}_m}{N_\Phi} \sideset{}{'}\sum_{\bf q} e^{iq_x(X_{j_1}-X_{j_3})}
 e^{-{q^2\over 2}} L_m\left(q^2\right),
\end{align}
where $X_j=2\pi j/b$ for $j=1,2,\dots,N_{\Phi}$.
In the above, the primed summation over momentum excludes the ${\bf q}=(0,0)$ component, which is cancelled by the uniform positive background charge.
The primed Kronecker delta is defined such that $\delta'_{m,n}=1$ if $m=n$ modulo $N_{\Phi}$, and 0 otherwise.
Similarly, the bilayer QH Hamiltonian can be rewritten in second quantization by making appropriate changes for $V_{\bf q}$, which now depends on the layer indices as described in Eq.~\eqref{eq:H_BQH}.

The torus geometry is nothing but the rectangular geometry with periodic boundary condition. 
We have chosen the torus geometry since it is completely free from the ``flux shift'' issue, which plagues the spherical geometry. 
In this geometry, all competing states belonging to the same filling factor can be compared on an equal footing. 
Another important advantage is that the effects of anisotropy can be investigated in a natural setting. 
In other words, the anisotropy in the band mass~\cite{PhysRevB.85.165318} as well as the dielectric tensor~\cite{PhysRevB.86.035122} can be naturally implemented in the Hamiltonian.

The exact ground state of $H_{1{\rm LL}}$ is obtained in the torus geometry by comparing all the lowest energy states obtained in each sector of total pseudomomentum, ${\bf K}=(K_x, K_y) = (m, n)$ in units of $(\frac{2\pi}{a}, \frac{2\pi}{b})$, where $m$ and $n$ are integers between $0$ and $N-1$~\cite{PhysRevLett.55.2095}.  
It is shown later that, except for few sporadic parameter regimes, the ground state of $H_{1{\rm LL}}$ tends to occur in the same pseudomomentum sector, which is at ${\bf K}=(6,6)$ in the case of the $N=12$ system.
To investigate the possible connection, we take the ground state of $H_{\rm BQH}$ in the same ${\bf K}=(6,6)$ sector.

Note that we perform ED almost exclusively in the $N=12$ system.
Under the constraint that we are only interested in even numbers of particles, which are necessary for pairing, the reason for focusing on the $N=12$ system is as follows.
First, the $N=8$ system suffers from a severe finite-size artifact, which forces the ground state to form Wigner crystals with 4 electrons in each layer of the bilayer QH system~\cite{PhysRevB.69.045319}.
Second, the $N=10$ system is still somewhat too small to represent the paired QH state correctly, considering the fact that, at $N=10$, there is no clear sign for the threefold quasidegeneracy in the ground state energy that is believed to be one of the most important topological characteristics of the paired QH state~\cite{PhysRevLett.101.016807,PhysRevB.80.241311,PhysRevB.78.155308}. 
While the quasidegeneracy is only approximately observed, it is apparent that, at least, the $N=10$ system is somewhat different from the rest of larger systems with $N \geq 12$.  
Third, any finite-size systems with $N \geq12$ systems should be fine.
However, only the $N=12$ system is within the reach of our current computer hardware capacity. 
While the single-layer problem in the half-filled SLL does not create too much trouble, the bilayer problem in the half-filled LLL generates too large a Hilbert space to be diagonalized exactly.
Specifically, the numbers of basis states are roughly given by $(_{N_{\Phi}}\!C_{N/2})^2/N_{\Phi}N$ and $_{N_{\Phi}}\!C_{N}/N_{\Phi}N$ for the bilayer and single-layer systems, respectively. 
At $N=12$, the number of basis states for the bilayer system is about $6.3 \times 10^7$, while that for the single-layer system is just about $9.4 \times 10^3$.
By contrast, at $N=14$, the number of basis states for the bilayer system is about $3.5 \times 10^9$, which is out of reach, while that for the single-layer system is about $1.0 \times 10^5$, which can be diagonalized without problem.

\section{Trial states}
\label{sec:Trial_states}
 
As mentioned previously, one of the main motivations for this work is the exact equivalence~\cite{PhysRevB.46.9586} between the MR Pfaffian wave function~\cite{moore1991nonabelions} and the antisymmetrized Halperin (331) wave function~\cite{halperin1983theory} of the bilayer QH system, where the Halperin (331) state is valid at interlayer distance roughly equal to the magnetic length, say, $d/l_{\rm B} \simeq 1\mbox{--}2$~\cite{PhysRevB.39.1932}.

Specifically, the MR Pfaffian wave function is written as follows:
\begin{align}
\label{eq:Psi_MR}
\Psi_{\rm MR} = \prod_{i<j} (z_i - z_j)^2 {\rm Pf}\left(\frac{1}{z_{k} - z_{l}}\right) ,
\end{align}  
where ${\rm Pf}(M_{ij})={\cal A}(M_{12}M_{34}\cdots M_{N-1,N})$ with ${\cal A}$ being the antisymmetrization operator.
Meanwhile, the Halperin (331) wave function is written as follows:
\begin{align}
\label{eq:Psi_331}
\Psi_{331} =& \prod_{i_{\in \uparrow}<j_{\in \uparrow}}  (z_i - z_j)^3  
\prod_{k_{\in \downarrow}<l_{\in \downarrow}} (z_k - z_l)^3
\nonumber \\
&\times \prod_{m_{\in \uparrow},n_{\in \downarrow} } (z_m - z_n) ,
\end{align}
where the top and bottom layers are denoted by the pseudospin $\uparrow$ and $\downarrow$ indices, respectively.  
Here, $i_{\in\uparrow}$ means that the $i$-th electron belongs to the top layer, or the pseudospin of the $i$-th electron is $\uparrow$. 
It is important to note that $\Psi_{331}$ in the above is only the space part of the total wave function, which contains both the space and pseudospin components. 
It is with help of the Cauchy identity~\cite{Book07Jain} that the MR Pfaffian wave function is shown to be entirely equivalent to the antisymmetrized Halperin (331) wave function up to a constant normalization factor; 
\begin{align}
\label{eq:MR_A331_equivalence}
\Psi_{\rm MR} = {\cal A} \Psi_{331} ,
\end{align}
where it is emphasized that ${\cal A}$ antisymmetrizes only the spatial coordinates inside the Halperin (331) wave function.

Both the MR Pfaffian and the Halperin (331) states are of course trial states, which, albeit very good in particular parameter regimes, are only approximations to the true ground states.  
A main question is how good these approximations are in the important parameter regimes.
In particular, we are interested in the ground state in the half-filled SLL for the pure Coulomb interaction, or some slightly modified interactions nearby.
While there are various experimentally important factors affecting the nature of the ground state such as Landau level mixing, finite thickness, and so on, we are first and foremost interested in what happens at the pure Coulomb interaction.
Unfortunately, the MR Pfaffian state (or its PH conjugate, the anti-Pfaffian state) has a relatively low overlap with the true ground state at the pure Coulomb interaction. 
(The MR Pfaffian and anti-Pfaffian states have the identical overlap with the Coulomb ground state due to the PH symmetry.)
The overlap can be enhanced by modifying the Coulomb interaction in some appropriate manners~\cite{PhysRevLett.84.4685,PhysRevLett.106.116801,PhysRevLett.80.1505}.
One of the most natural manners is via increasing the $m=1$ Haldane pseudopotential in Eq.~\eqref{eq:delta_V_1} up to $\delta V^{(1)}_1/V^{(1)}_1 \simeq 5 \mbox{--} 10 \%$. 
For convenience, let us denote the exact ground state in the half-filled SLL as $\Psi_{5/2}[\delta V^{(1)}_1]$, where it is explicitly shown that $\Psi_{5/2}$ is a function of $\delta V^{(1)}_1$.

As mentioned previously, the Halperin (331) state is known to be a good trial state for the bilayer QH system at $d/l_{\rm B} \simeq 1 \mbox{--} 2$~\cite{PhysRevB.39.1932,PhysRevB.64.085313}.
In fact, the bilayer QH system exhibits a series of quantum phase transitions as a function of interlayer distance~\cite{Book07Jain,PhysRevB.64.085313}.  
Specifically, at $d/l_{\rm B} \ll 1$, the ground state is well approximated by the pseudospin-unpolarized CF sea state at half filling, $\Psi_{^2{\rm CFS}|_{S=0}}$, where $^2$CFS stands for the Fermi sea state of composite fermions capturing two vortices and $S=0$ means that the total pseudospin is zero.  
It is interesting to note that, in the absence of (pseudospin) Zeeman coupling, the CF sea is completely unpolarized despite the na\"{i}ve expectation from the Hund's rule~\cite{PhysRevLett.80.4237}.
For the later use, let us denote the fully pseudospin-polarized CF sea state at half filling as  
$\Psi_{^2{\rm CFS}|_{S_{\rm max}}}$ in comparison with $\Psi_{^2{\rm CFS}|_{S=0}}$.
Meanwhile, at $d/l_{\rm B} \gg 1$, the ground state is well approximated by the product state of two CF seas at quarter filling, $\Psi_{^4{\rm CFS}_\uparrow}\otimes \Psi_{^4{\rm CFS}_\downarrow}$, where $^4$CFS$_\uparrow$ and $^4$CFS$_\downarrow$ denote the Fermi sea state of pseudospin-up and down composite fermions capturing four vortices, respectively.

To summarize, the ground state of the bilayer QH system undergoes a sequence of two quantum phase transitions, which can be roughly understood in terms of the phase transitions from $\Psi_{^2{\rm CFS}|_{S=0}}$, to $\Psi_{331}$, and to $\Psi_{\rm ^4CFS_\uparrow}\otimes \Psi_{\rm ^4CFS_\downarrow}$ as an increasing function of $d/l_{\rm B}$~\cite{PhysRevB.64.085313}.
Specifically, denoting the true ground state of the bilayer QH system as $\Psi_{\rm BQH}[d]$, this means that  
\begin{align}
\label{eq:Psi_BQH}
\Psi_{\rm BQH}[d] \simeq
\begin{cases}
\Psi_{^2{\rm CFS}|_{S=0}} & d/l_{\rm B} \ll 1 \\
\Psi_{331}                        & d/l_{\rm B} \simeq 1 \mbox{--} 2 \\
\Psi_{^4{\rm CFS}_\uparrow}\otimes \Psi_{^4{\rm CFS}_\downarrow} & d/l_{\rm B} \gg 1 
\end{cases} .
\end{align}

As mentioned in Eq.~\eqref{eq:MR_A331_equivalence}, the MR Pfaffian state, $\Psi_{\rm MR}$, is entirely equivalent to the antisymmetrized Halperin (331) state, ${\cal A} \Psi_{331}$.
However, neither $\Psi_{\rm MR}$ nor $\Psi_{331}$ are exactly identical to the true ground state in the SLL or that of the bilayer QH system, respectively. 
Therefore, it is interesting to ask how the exact ground state of the SLL system, $\Psi_{5/2}[\delta V^{(1)}_1]$, is actually related with the antisymmetrized exact ground state of the bilayer QH system, ${\cal A}\Psi_{\rm BQH}[d]$.
Formally, we are interested in the square of overlap between the two states defined as follows:
\begin{align}
|{\cal O}|^2 = |\langle  {\cal A} \Psi_{\rm BQH}[d]  | \Psi_{5/2}[\delta V^{(1)}_1] \rangle|^2 ,
\end{align}
which can be computed as a function of $\delta V^{(1)}_1$ and $d$.
Physically speaking, 
we ask if the ground state in the SLL can be in general mapped onto the antisymmetrized ground state of the bilayer QH system.  
If so, such a bilayer mapping can reveal the underlying bipartite structure, which is instrumental to pairing.

In a more practical view, based on the fact that ${\cal A} \Psi_{\rm BQH}[d]$ interpolates between the unpaired CF sea state at $d/l_{\rm B} \ll 1$ and the paired MR Pfaffian state at $d/l_{\rm B} \simeq 1\mbox{--} 2$, 
$d/l_{\rm B}$ can be regarded as a good variational parameter that can tune the ``strength'' of pairing. 
As mentioned in Sec.~\ref{sec:Introduction}, M\"{o}ller and Simon~\cite{PhysRevB.77.075319} have previously proposed a trial wave function scheme for the BCS paired CF states, which can be adiabatically connected to the MR Pfaffian state via careful tuning of multiple variational parameters. 
One obvious advantage of our work is that we have only a single variational parameter, $d/l_{\rm B}$, which provides a natural interpolation scheme without any detailed tuning processes.

In addition to providing a natural interpolation between the CF sea and the MR Pfaffian state, our scheme presents a very intriguing question that was not considered before; 
what is the nature of the antisymmetrized bilayer ground state at $d/l_{\rm B} \gg 1$? 
It is shown in the following sections that this question is actually very important since, at the pure Coulomb interaction, i.e., $\delta V^{(1)}_1=0$, 
$|\langle \Psi_{5/2}[\delta V^{(1)}_1] | {\cal A} \Psi_{\rm BQH}[d] \rangle|^2$ is maximal at $d/l_{\rm B} \gg 1$, not at $d/l_{\rm B} \simeq 1.5\mbox{--}2$, where the MR Pfaffian state is relevant. 
We discuss the physical meaning of the antisymmetrized bilayer ground state at $d/l_{\rm B} \gg 1$ in detail in the following sections.

\section{Antisymmetrized bilayer ground state}
\label{sec:ABGS}

\subsection{Antisymmetrization in second quantization}
\label{sec:ASQ}

Before discussing the results obtained from the antisymmetrized bilayer ground state, we first explain how to apply the antisymmetrization operator, ${\cal A}$, onto the exact bilayer ground state, $\Psi_{\rm BQH}[d]$ that is obtained via ED in second quantization. 
While rarely performed in practice, applying the antisymmetrization operator to a given bilayer wave function is straightforward in concept, at least, when the mathematical form of the bilayer wave function is given analytically. 
That is to say, the antisymmetrization operator swaps the spatial coordinates of pseudospin-up electrons with those of pseudospin-down electrons in all possible permutations, attach appropriate signs depending on the parity of the permutation, and finally add all thus-generated wave functions.  
The situation becomes complicated when the analytic form of the bilayer wave function is not known, which is particularly the case for the ground state obtained via ED in second quantization.
Applying the antisymmetrization operator to the exact bilayer ground state has not been performed before to the best of our knowledge. 

When obtained via ED, the ground state wave function is represented by a set of numbers, which are the amplitudes of corresponding basis states.
Each basis state is in turn represented by a sequence of ``single-site quantum states'' at individual Landau-gauge orbitals, which themselves are enumerated by their momentum indices.
There are four possible ``single-site quantum states'' at a given Landau-gauge orbital, which are the empty, pseudospin-up, pseudospin-down, and doubly occupied states.
For convenience, let us denote them by $|0\rangle$, $\left|\uparrow\right\rangle$, $\left|\downarrow\right\rangle$, and $|2\rangle$, respectively.

Now, let us think about what operation the antisymmetrization operator performs on the basis states. 
The most important operation is to remove all doubly occupied states.  
In other words, any states containing the doubly occupied states, for example, $\left|\cdots \uparrow~0~\downarrow~\uparrow~2~\downarrow\cdots\right\rangle$, are to be annihilated upon the application of the antisymmetrization operator.
Next, since the pseudospin index is no longer meaningful after antisymmetrization, both $\left|\uparrow\right\rangle$ and $\left|\downarrow\right\rangle$ are mapped onto the singly occupied state, $|1\rangle$~\cite{PhysRevLett.101.036804}. 
Consequently, $|0\rangle$ and $|1\rangle$ consist of the entire Hilbert space at each orbital after antisymmetrization. 
Finally, the amplitudes of the surviving basis states are rescaled so that the total wave function is normalized, which finally gives rise to ${\cal A} \Psi_{\rm BQH}[d]$.

Another way of looking at the antisymmetrization operator is that it is the projection operator that projects a given state onto the fully pseudospin-polarized sector with pseudospin aligned, say, along the $x$ direction. 
In this view, $\left|\uparrow\right\rangle$ and $\left|\downarrow\right\rangle$ can be understood as $\frac{1}{\sqrt{2}}(\left| \rightarrow\right\rangle +  \left|\leftarrow\right\rangle  )$ and $\frac{1}{\sqrt{2}}(\left| \rightarrow\right\rangle -  \left|\leftarrow\right\rangle  )$, respectively, where  $\left| \rightarrow\right\rangle$ and $\left| \leftarrow\right\rangle$ denote the eigenstates of the pseudospin along the $x$ direction with positive and negative eigenvalue, respectively.  
Keeping only the $\left| \rightarrow\right\rangle$ component, the antisymmetrization operator simply projects both $\left|\uparrow\right\rangle$ and $\left|\downarrow\right\rangle$ onto $\left| \rightarrow\right\rangle$, which can be alternatively denoted as $|1\rangle$ as done so in the above.  
The doubly occupied state is naturally annihilated by the antisymmetrization operator due to the Pauli exclusion principle. 


In what follows, we check the validity of the antisymmetrization procedure described in the above by making comparison between ${\cal A} \Psi_{331}$ with $\Psi_{\rm MR}$, which should be exactly identical.
Then, we compute the overlap between ${\cal A} \Psi_{\rm BQH}[d]$ and three antisymmetrized trial states of the bilayer QH system, which are given by $\Psi_{^2{\rm CFS}|_{S_{\rm max}}}$, $\Psi_{\rm MR}$, and ${\cal A} \Psi_{^4{\rm CFS}_\uparrow}\otimes\Psi_{^4{\rm CFS}_\downarrow}$.

\subsection{Comparison with trial states}
\label{sec:Comparison}

The Halperin (331) state can be obtained exactly by constructing the model Hamiltonian with an appropriately chosen set of the Haldane pseudopotentials.  
Specifically, the Halperin (331) state can be obtained as the exact zero-energy ground state of the following Hamiltonian~\cite{PhysRevB.34.2670,PhysRevB.39.1932,Book87Prange}:
\begin{align}
\label{eq:H_331}
H_{331} = \sum_{m=0}^{\infty}{2 \over N_{\Phi}}\sum_{{\bf q}}e^{-{q^2/2}} L_m(q^2)\sum_{i<j}  V^{\sigma_i,\sigma_j}_m e^{i{\bf q}\cdot ({\bf R}_{i}-{\bf R}_{j})},
\end{align}
where $V^{\sigma_i,\sigma_j}_m = \delta_{m,1}$ for the interaction between electrons with the same pseudospin, and $\delta_{m,0}$ for the interaction between electrons with different pseudospins.
There are eight degenerate zero-energy copies of the Halperin $(331)$ state in the torus geometry, which occur at pseudomomentum ${\bf K}=(0,0)$, $(N/2, 0)$, $(0, N/2)$, $(N/2, N/2)$, and those due to the center-of-mass degeneracy~\footnote{E. Keski-Vakkuri and X.-G. Wen, Int. J. Mod. Phys. B {\bf 7}, 4227 (1993)].}.

Meanwhile, the MR Pfaffian state can be obtained as the exact zero-energy ground state of the following three-body interaction Hamiltonian~\cite{PhysRevLett.66.3205,greiter1992paired}:
\begin{align}
\label{eq:H_MR}
H_{\rm MR}=-\!\sum_{i<j<k}\!{\cal S}_{ijk}\left(\nabla_i^4\nabla_j^2\right)\delta^2({\bf r}_i-{\bf r}_j)\delta^2({\bf r}_j-{\bf r}_k),
\end{align}
where ${\cal S}_{ijk}$ is the symmetrization operator among the $(i,j,k)$ indices. 
There are six degenerate zero-energy copies of the MR Pfaffian state in the torus geometry, which occur at pseudomomentum ${\bf K}=(N/2, 0)$, $(0, N/2)$, $(N/2, N/2)$, and those due to the center-of-mass degeneracy~\cite{greiter1992paired}.

To confirm the validity of the antisymmetrization procedure discussed in the preceding section, we have performed ED of $H_{331}$ and $H_{\rm MR}$ in various finite-size systems, which has shown that the antisymmetrized ground state of $H_{331}$ is indeed exactly identical to the ground state of $H_{\rm MR}$ within numerical accuracy.

Now, we investigate how the antisymmetrized exact ground state of the bilayer QH system, ${\cal A} \Psi_{\rm BQH}[d]$, obtained via ED is related with the following three trial states; $\Psi_{^2{\rm CFS}|_{S_{\rm max}}}$, $\Psi_{\rm MR}$, and ${\cal A} \Psi_{^4{\rm CFS}_\uparrow}\otimes\Psi_{^4{\rm CFS}_\downarrow}$.  
These three trial states are obtained from the following conjecture:
\begin{align}
\label{eq:A_Psi_BQH}
{\cal A}\Psi_{\rm BQH}[d] \simeq
\begin{cases}
\Psi_{^2{\rm CFS}|_{S_{\rm max}}} & d/l_{\rm B} \ll 1 \\
\Psi_{\rm MR}                        & d/l_{\rm B} \simeq 1 \mbox{--} 2 \\
{\cal A}\Psi_{^4{\rm CFS}_\uparrow}\otimes \Psi_{^4{\rm CFS}_\downarrow} & d/l_{\rm B} \gg 1 
\end{cases} ,
\end{align}
which is related with Eq.~\eqref{eq:Psi_BQH}, where the bilayer ground state (before antisymmetrization) is approximated by the three trial states of 
$\Psi_{^2{\rm CFS}|_{S=0}}$, $\Psi_{331}$, and $\Psi_{^4{\rm CFS}_\uparrow}\otimes \Psi_{^4{\rm CFS}_\downarrow}$
as an increasing function of $d/l_{\rm B}$.

Let us examine the physical meaning of Eq.~\eqref{eq:A_Psi_BQH} for a moment.
First, it is easy to understand the second line since $\Psi_{\rm BQH}[d \simeq 1\mbox{--}2]\simeq \Psi_{331}$ and $\Psi_{\rm MR}={\cal A}\Psi_{331}$.
The third line is simple in the sense that, at $d/l_{\rm B} \gg 1$, each layer of the bilayer QH system becomes completely independent, in which situation the pseudospin-up and down composite fermions form the Fermi sea states of their own at quarter filling.  
We obtain ${\cal A} \Psi_{^4{\rm CFS}_\uparrow}\otimes \Psi_{^4{\rm CFS}_\downarrow}$ by antisymmetrizing the ground state of the bilayer QH system at sufficiently large interlayer distance, which is known to be well described by $\Psi_{^4{\rm CFS}_\uparrow}\otimes \Psi_{^4{\rm CFS}_\downarrow}$~\cite{PhysRevB.64.085313}.

One of the most interesting predictions in Eq.~\eqref{eq:A_Psi_BQH} is the first line where the trial state is given by the fully pseudospin-polarized CF sea state, not the unpolarized counterpart.  
The reason is as follows.
At $d=0$, the bilayer QH problem simply reduces to the single-layer problem with real spin degree of freedom without Zeeman coupling. 
In this situation, the ground state is known to be the CF sea state with the unpolarized spin configuration~\cite{PhysRevLett.80.4237}.
Now, at $d/l_{\rm B} \ll 1$, but not exactly equal to 0, the spin rotational symmetry is slightly broken so that ground state can, in general, have nonzero components in various total spin sectors. 
Meanwhile, since the interaction between electrons is only slightly modified, the overall structure of the ground state is predicted to still remain as the CF sea state.
Combining these two observations, one can conjecture that, at $d/l_{\rm B} \ll 1$,
\begin{align} 
\label{eq:2composite fermions_conjecture}
\Psi_{\rm BQH}[d] \simeq \sum_{S=0}^{S_{\rm max}} \lambda _S \Psi_{^2{\rm CFS}|_{S}} ,
\end{align}
where $\lambda_S$ is the relative weight in the sector of total spin $S$. 
Based on the previously-mentioned fact that the single-layer ground state is the CF sea state with the unpolarized spin configuration, $\lambda_S$ is presumed to be peaked around $S=0$ as $d/l_{\rm B}$ decreases. 
Now, considering the fact that the antisymmetrization operator projects a given bilayer state onto the fully pseudospin-polarized sector, we arrive at the prediction that the antisymmetrized bilayer ground state should be well described by $\Psi_{^2{\rm CFS}|_{S_{\rm max}}}$ at $d/l_{\rm B} \ll 1$.
It is shown below that this prediction is actually correct.

\begin{figure}[]
\includegraphics[width=1\columnwidth]{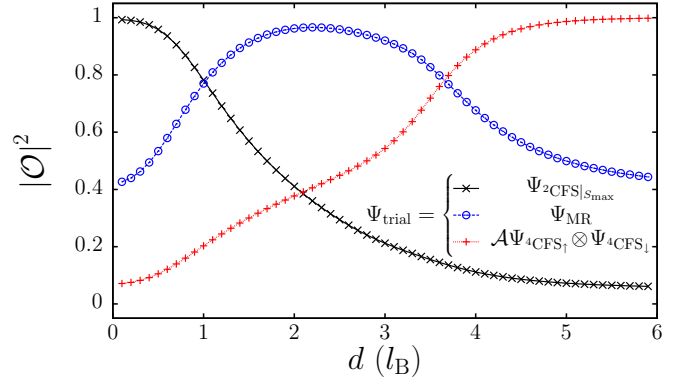}
\caption{
Square of overlap, $|{\cal O}|^2$, between the antisymmetrized ground state of the bilayer QH system, ${\cal A}\Psi_{\rm BQH}[d]$, and three trial states, $\Psi_{\rm trial}$; (i) the fully pseudospin-polarized CF sea state at half filling, $\Psi_{^2{\rm CFS}|_{S_{\rm max}}}$, (ii) the MR Pfaffian state, $\Psi_{\rm MR}$, and (iii) the antisymmetrized product state of two CF seas at quarter filling with pseudospin up and down, 
${\cal A}\Psi_{^4{\rm CFS}_\uparrow}\otimes \Psi_{^4{\rm CFS}_\downarrow}$.
Here, $\Psi_{\rm BQH}[d]$ is obtained via ED of the bilayer QH system in the torus geometry with the total number of electrons $N=12$ and the number of pseudospin up and down electrons $N_\uparrow=N_\downarrow=6$. 
The aspect ratio of the rectangular unit call is set as $r_a=0.98$.
All calculations are performed at pseudomomentum ${\bf K}=(6,6)$. 
Note that $\Psi_{^4{\rm CFS}_\uparrow}\otimes \Psi_{^4{\rm CFS}_\downarrow}$ is obtained as the exact ground state of the bilayer QH system at sufficiently large interlayer distance, say, $d/l_{\rm B}=10$. 
}
\label{fig:Comparison}
\end{figure}

Figure~\ref{fig:Comparison} shows the square of overlap between ${\cal A} \Psi_{\rm BQH}[d]$ and the three trial states of $\Psi_{^2{\rm CFS}|_{S_{\rm max}}}$, $\Psi_{\rm MR}$, and ${\cal A} \Psi_{^4{\rm CFS}_\uparrow}\otimes\Psi_{^4{\rm CFS}_\downarrow}$ in the $N=12$ system as a function of $d/l_{\rm B}$. 
The behavior of the overlap is entirely consistent with what is expected, reinforcing the fact that the three trial states are good approximations in their respective parameter regimes.

First, as expected in the above, the square of overlap between ${\cal A} \Psi_{\rm BQH}[d]$ and $\Psi_{^2{\rm CFS}|_{S_{\rm max}}}$ is close to unity at $d/l_{\rm B} \ll 1$ and decreases monotonically as $d/l_{\rm B}$ increases.  
To test the conjecture described in Eq.~\eqref{eq:2composite fermions_conjecture}, 
we have computed the so-called survival ratio, $R_{\rm s}$, which is defined as follows: 
\begin{align}
\label{eq:r_s}
R_{\rm s} = \left|\left\langle \Psi_{\rm BQH}[d] \left| {\cal P}_{S_z=0} {\cal R}_y (\pi/2) \right|  {\cal A} \Psi_{\rm BQH} [d] \right\rangle\right|^2 ,
\end{align}
where ${\cal R}_y(\pi/2)$ is the pseudospin rotation operator with respect to the $y$ axis by $\pi/2$ and
${\cal P}_{S_z=0}$ is the projection operator onto the $S_z=0$ sector, where the number of pseudospin-up electrons is the same as that of pseudospin-down electrons.
Note that ${\cal R}_y(\pi/2)$ is necessary since ${\cal A} \Psi_{\rm BQH} [d]$ is in the fully pseudospin-polarized sector, whose pseudospin direction is not specified. 
Assuming that the original pseudospin direction of ${\cal A} \Psi_{\rm BQH} [d]$ is along the $z$ direction, i.e., all electrons are in the same layer, we have to rotate its total pseudospin with respect to the $y$ direction by $\pi/2$ to construct the fully pseudospin-polarized state along the $x$ direction, which, on average, has the equal number of pseudospin-up and down electrons.  
Then, we further project the rotated state onto the $S_z=0$ sector, where the number of pseudospin-up electrons is fixed to be precisely the same as that of pseudospin-down electrons.
Finally, we compute its overlap with $\Psi_{\rm BQH}[d]$ to obtain $R_{\rm s}$ that measures how much portion of $\Psi_{\rm BQH} [d]$ survives after antisymmetrization.
Results from ED reveal that $R_{\rm s}$ vanishes completely in the limit of $d/l_{\rm B} \ll 1$, which shows that, at $d/l_{\rm B} \ll 1$, the antisymmetrization operator annihilates almost the entire components of $\Psi_{\rm BQH}[d]$ except for a very small portion belonging to the fully pseudospin-polarized sector, which itself, upon renormalization, has the close-to-unity overlap with $\Psi_{^2{\rm CFS}|_{S_{\rm max}}}$.

Next, it is observed in Fig.~\ref{fig:Comparison} that the square of overlap between ${\cal A} \Psi_{\rm BQH}[d]$ and $\Psi_{\rm MR}$ is broadly peaked at $d/l_{\rm B} \simeq 2$ with its maximum value being close to unity, which is completely consistent with our prediction. 
Quantitatively, the square of overlap, $\left| \left\langle{\cal A} \Psi_{\rm BQH}[d] | \Psi_{\rm MR} \right\rangle \right|^2$, becomes about $97 \%$ at $d/l_{\rm B} \simeq 2.2$.
Despite this complete consistency at $d/l_{\rm B} \simeq 2$, we observe a somewhat surprising fact that $\left| \left\langle{\cal A} \Psi_{\rm BQH}[d] | \Psi_{\rm MR} \right\rangle \right|^2$ remains large ($\simeq 40 \%$) even in the limit of both small and large $d/l_{\rm B}$.
This suggests that substantial pairing correlations exist in both limits.

The existence of the pairing correlation at small $d/l_{\rm B}$ is consistent with the previous result by Rezayi and Haldane~\cite{PhysRevLett.84.4685}.
Based on ED, it is concluded by the authors that the system at the half-filled second Landau level may {\it always} be paired and smoothly crosses over from a weakly paired regime described by the CF sea state to a strongly paired regime by the MR Pfaffian state.
This conclusion is consistent with our finding that the CF sea state, i.e., ${\cal A}\Psi_{\rm BQH}[d/l_{\rm B} \ll 1]$, has a sizable overlap with the MR Pfaffian state.

The possible existence of pairing correlation at large $d/l_{\rm B}$ has not been studied before to the best of our knowledge.
At first, this possibility seems rather unlikely since the product state of two CF seas at quarter filling, $\Psi_{^4{\rm CFS}_\uparrow}\otimes \Psi_{^4{\rm CFS}_\downarrow}$, before antisymmetrization does not have any resemblance to an incompressible state, let alone a paired state. 
A surprise occurs after antisymmetrization, as shown via the overlap between the bilayer ground and trial states before and after antisymmetrization. 
Specifically, $\left| \langle \Psi_{\rm BQH}[d] | \Psi_{331} \rangle \right|^2$ is less than $1\%$, while $\left| \langle {\cal A} \Psi_{\rm BQH}[d] | \Psi_{\rm MR} \rangle \right|^2$ is over $40\%$ at $d/l_{\rm B} \gtrsim 6$.
Note that $\Psi_{\rm BQH}[d]$ is essentially given by $\Psi_{^4{\rm CFS}_\uparrow}\otimes \Psi_{^4{\rm CFS}_\downarrow}$ for sufficiently large $d/l_{\rm B}$, say, $\gtrsim 6$~\cite{PhysRevB.64.085313}.

A possible conclusion so far from Fig.~\ref{fig:Comparison} is that the antisymmetrized bilayer ground state makes a transition or smooth crossover from a weakly paired state described by the CF sea state at half filling, $\Psi_{^2{\rm CFS}|_{S_{\rm max}}}$ to a strongly paired state by the MR state, $\Psi_{\rm MR}$, which is consistent with the previous result by Rezayi and Haldane~\cite{PhysRevLett.84.4685}.
At large $d/l_{\rm B}$, the antisymmetrized bilayer ground state reduces to the antisymmetrized product state of two CF seas at quarter filling, ${\cal A}\Psi_{^4{\rm CFS}_\uparrow}\otimes \Psi_{^4{\rm CFS}_\downarrow}$, which might also contain substantial pairing correlation. 
We believe that ${\cal A}\Psi_{^4{\rm CFS}_\uparrow}\otimes \Psi_{^4{\rm CFS}_\downarrow}$ can be as strongly paired as, if not more so than, $\Psi_{\rm MR}$ in the sense that the ``pairing strength'' is related with the discrepancy between the intralayer and interlayer interaction between electrons in the bilayer QH system.
Note that, the larger  $d/l_{\rm B}$ gets, the stronger the discrepancy becomes. 
In other words, the interaction between electrons (or quasiparticles) in different layers becomes less repulsive than that within the same layer.     
Relatively speaking, electrons could be regarded as being paired between different layers.  
If so, the pairing strength becomes stronger, as $d/l_{\rm B}$ increases.

In the following section, we provide more conclusive evidence supporting the above conclusion by computing the overlap between the antisymmetrized bilayer ground state, ${\cal A }\Psi_{\rm BQH}[d]$, and the exact ground state in the half-filled SLL, $\Psi_{5/2}[\delta V^{(1)}_1]$, as a function of both $d/l_{\rm B}$ and $\delta V^{(1)}_1$.
Specifically, it is shown in the following section that ${\cal A} \Psi_{\rm BQH}[d]$ is intimately connected with $\Psi_{5/2}[\delta V^{(1)}_1]$ along a certain trajectory in the parameter space of $(d/l_{\rm B}, \delta V^{(1)}_1/V^{(1)}_1)$, which generates high overlaps.  
We call the one-to-one relationship along this trajectory the bilayer mapping.


\section{Bilayer mapping}
\label{sec:Bilayer_mapping}

\begin{figure*}[t]
\includegraphics[width=2\columnwidth]{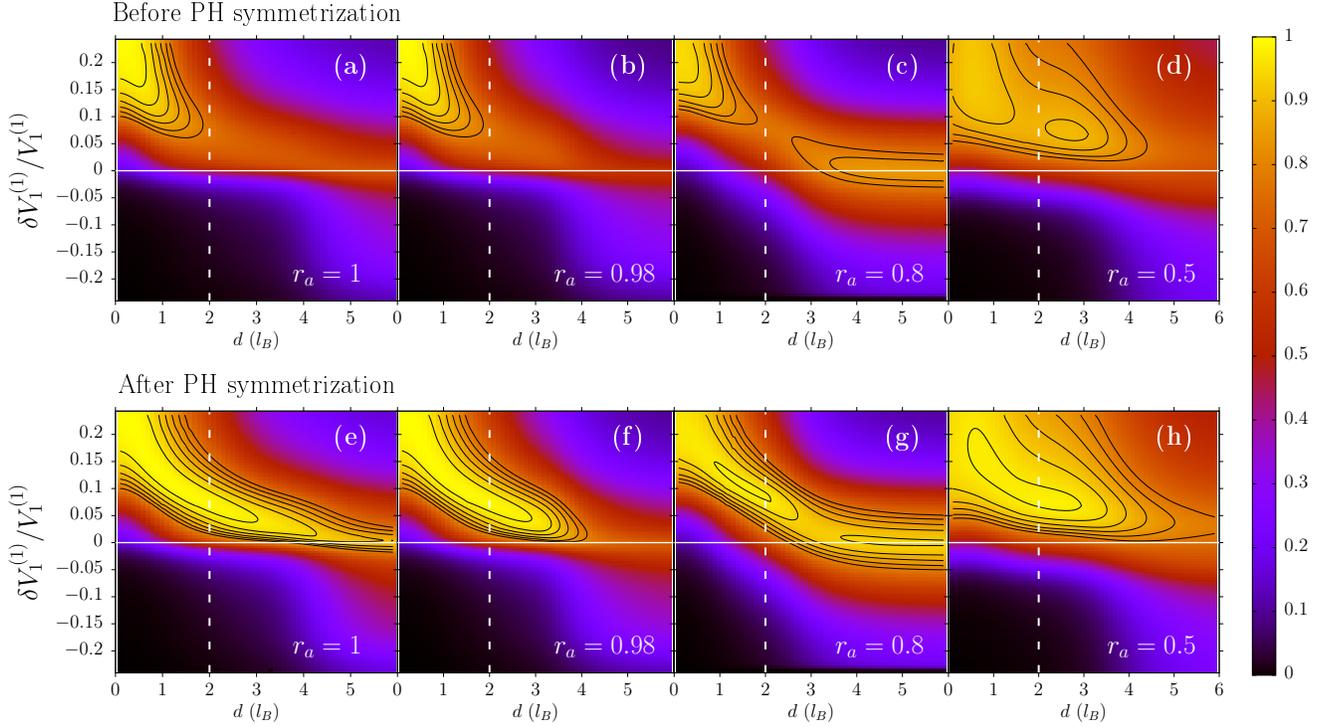}
\caption{
Square of overlap [(a)\mbox{--}(d)] between the exact $5/2$ state, $\Psi_{5/2}[\delta V^{(1)}_1]$, and
the antisymmetrized bilayer ground state, ${\cal A}\Psi_{\rm BQH}[d]$, as a function of $d/l_{\rm B}$ and $\delta V^{(1)}_1/V^{(1)}_1$.
Here, $\Psi_{5/2}[\delta V^{(1)}_1]$ is obtained via ED of the SLL Hamiltonian in Eq.~\eqref{eq:H_LLn_pseudo}, $H_{1{\rm LL}}$, in the $N=12$ system.
Meanwhile, ${\cal A}\Psi_{\rm BQH}[d]$ is obtained by applying the antisymmetrization operator onto the exact ground state of the BQH system, $\Psi_{\rm BQH}[d]$, which is in turn obtained via ED of the bilayer QH Hamiltonian, $H_{\rm BQH}$, in Eq.~\eqref{eq:H_BQH} in the $N=12$ system with $N_\uparrow=N_\downarrow=6$.
While the overall pattern is similar, the square of overlap 
is improved substantially by applying the PH symmetrization operator, ${\cal S}_{\rm PH}$, onto ${\cal A}\Psi_{\rm BQH}[d]$, the results of which are shown in (e)\mbox{--}(h).
For convenience, several contour lines are drawn, along which the square of overlap takes the constant values of $77\%$, $81\%$, $85\%$, $89\%$, $93\%$, and $97 \%$ from low to high values.
All calculations are performed at pseudomomentum ${\bf K}=(6,6)$ in the torus geometry.
The aspect ratio, $r_a$, of the rectangular unit cell is varied from 1.0 [(a) and (e)], to 0.98 [(b) and (f)], to 0.8 [(c) and (g)], and to 0.5 [(d) and (h)].
Note that, among all pseudomomentum sectors, the global ground state of $H_{\rm 1LL}$ occurs at ${\bf K}=(6,6)$ in almost the entire range of $\delta V^{(1)}_1$ except for few sporadic windows regardless of $r_a$. 
Even when it does not, the global ground state tends to occur at one of the three quasidegenerate sectors among ${\bf K}=(6,6)$, $(6,0)$, and $(0,6)$.
The quasidegeneracy is most pronounced in the regime of $0 \lesssim \delta V^{(1)}_1/V^{(1)}_1 \lesssim 0.1$. 
The path following $\delta V^{(1)}_1=0$ is denoted by white solid lines, where the interaction is at the pure Coulomb point.
The path following $d/l_{\rm B} =2$ is denoted by  white dashed lines, where ${\cal A}\Psi_{\rm BQH}[d] \simeq \Psi_{\rm MR}$.
%
}
\label{fig:Phase_diagram}
\end{figure*}

In this section, we show that the exact $5/2$ state, $\Psi_{5/2}[\delta V^{(1)}_1]$, is intimately connected with the antisymmetrized bilayer ground state, ${\cal A} \Psi_{\rm BQH}[d]$, by computing the square of overlap between the two states, $| \langle {\cal A} \Psi_{\rm BQH}[d] | \Psi_{5/2}[\delta V^{(1)}_1] \rangle |^2$, which is shown in Fig.~\ref{fig:Phase_diagram} as a function of $d/l_{\rm B}$ and $\delta V^{(1)}_1/V^{(1)}_1$.
Note that $\Psi_{5/2}[\delta V^{(1)}_1]$ is obtained via ED of the 1LL Hamiltonian in Eq.~\eqref{eq:H_LLn_pseudo}, $H_{1{\rm LL}}$, at half filling.
Meanwhile, ${\cal A}\Psi_{\rm BQH}[d]$ is obtained by applying the antisymmetrization operator onto the exact ground state of the bilayer QH system, $\Psi_{\rm BQH}[d]$ that is in turn obtained via ED of the bilayer QH Hamiltonian in Eq.~\eqref{eq:H_BQH}, $H_{\rm BQH}$, at half filling.

Following Rezayi and Haldane~\cite{PhysRevLett.84.4685}, ${\cal A}\Psi_{\rm BQH}[d]$ can be further improved by applying the PH symmetrization operator, ${\cal S}_{\rm PH}$, onto ${\cal A}\Psi_{\rm BQH}[d]$, which restores the exact PH symmetry of the two-body interaction Hamiltonian in the half-filled Landau level.
For convenience, let us denote the PH-symmetry-restored antisymmetrized bilayer ground state as ${\cal S}_{\rm PH} \{ {\cal A} \Psi_{\rm BQH}[d] \}$.
See Appendix~\ref{appendix:PH_symmetrization} for details regarding how to implement PH symmetrization in the torus geometry.
It is interesting to note that the antisymmetrization operation is not necessarily compatible with the PH symmetry, as revealed by the correspondence, ${\cal A}\Psi_{331}=\Psi_{\rm MR}$, where $\Psi_{331}$ is PH symmetric, while $\Psi_{\rm MR}$ is not.
Figure~\ref{fig:Phase_diagram} shows the square of overlap both before [(a)\mbox{--}(d)] and after [(e)\mbox{--}(h)] PH symmetrization. 
It is shown in Fig.~\ref{fig:Phase_diagram} that, regardless of PH symmetrization, the overall pattern of the overlap is similar, while its quantitative value is enhanced substantially after its application.

One of the most prominent features in Fig.~\ref{fig:Phase_diagram} is that there is a clear trajectory (colored in yellow) in the parameter phase space, along which the square of overlap is high and close to unity, being especially so after PH symmetrization.
Overall, the high-overlap trajectory connects the CF sea state occurring at $\delta V^{(1)}_1/V^{(1)}_1 \gg 1$ and $d/l_{\rm B} \simeq 0$ with the MR Pfaffian state occurring at $0< \delta V^{(1)}_1/V^{(1)}_1 \lesssim 0.1$ and $d/l_{\rm B} \simeq 2$, and finally with the antisymmetrized product state of two CF seas at quarter filling occurring at $\delta V^{(1)}_1/V^{(1)}_1 =0$ and $d/l_{\rm B} \gg 1$. 
This high-overlap trajectory defines the bilayer mapping between the exact $5/2$ state and the antisymmetrized bilayer ground state.

To be specific, let us elaborate on the behavior of the square of overlap at unity aspect ratio, i.e., $r_a=1$ in Fig.~\ref{fig:Phase_diagram}~(a) and (e).
First, at $d/l_{\rm B} \rightarrow 0$, the square of overlap becomes over $99 \%$ at $\delta V^{(1)}_1/V^{(1)}_1 \simeq 0.25$ and approaches unity as $\delta V^{(1)}_1/V^{(1)}_1$ increases further.
This behavior is obtained regardless of PH symmetrization.
Second, at $d/l_{\rm B} \simeq 2$, the square of overlap reaches the maximum of about $75 \%$ at $\delta V^{(1)}_1/V^{(1)}_1 \simeq 0.07$ before PH symmetrization.
After PH symmetrization, the square of overlap is enhanced substantially to reach the maximum of $99 \%$ around the same value of $\delta V^{(1)}_1/V^{(1)}_1$.
Finally, being more or less saturated at sufficiently large $d/l_{\rm B}$,
the square of overlap reaches the maximum of $60\mbox{--}70 \%$ around $\delta V^{(1)}_1/V^{(1)}_1 \simeq 0$ before PH symmetrization. 
After PH symmetrization, the maximum value is again enhanced substantially to become more than $90 \%$.

Our results at $r_a=1$ are fully consistent with the previous results obtained by Rezayi and Haldane for the $N=10$ system~\cite{PhysRevLett.84.4685}. 
To see this, first, note that the MR Pfaffian state and the CF sea state are obtained along the paths following $d/l_{\rm B} \simeq 2$ (denoted via white dashed lines in the figure) and $d/l_{\rm B}\simeq 0$, respectively.  
Along the path of $d/l_{\rm B} \simeq 2$, where the MR Pfaffian state is valid, the overlap is high within the window of $0 < \delta V^{(1)}_1/V^{(1)}_1 \lesssim 0.1$, which is consistent with the previous result. 
In particular, the square of overlap becomes close to unity after PH symmetrization. 
Similarly, along the path of $d/l_{\rm B} \simeq 0$, where the CF sea state is valid, the overlap approaches unity in the limit of $\delta V^{(1)}_1/V^{(1)}_1 \gg 1$, which is again consistent.
Incidentally, by using the same computer code used for the $N=12$ system, we have performed ED of the $N=10$ system, which reproduces exactly the same results as those obtained by Rezayi and Haldane~\cite{PhysRevLett.84.4685}.

One of the most important discoveries in this work is that the exact $5/2$ state at the Coulomb point, $\delta V^{(1)}_1=0$, has the maximal overlap with the antisymmetrized bilayer ground state at $d/l_{\rm B} \gg 1$, not at $d/l_{\rm B} \simeq 1 \mbox{--} 2$, where the antisymmetrized bilayer ground state is given by the MR Pfaffian state.
While it is consistent with our expectation in Sec.~\ref{sec:Trial_states} that the antisymmetrized bilayer ground state at $d/l_{\rm B} \gg 1$ may have substantial pairing correlation, this result is quite surprising since na\"{i}vely the paired QH state is expected to be connected with an incompressible state in the bilayer QH system. 
Since there is only one candidate for the incompressible state in the bilayer QH system, which is the Halperin (331) state, a natural expectation is that the exact $5/2$ state is connected with the antisymmetrized bilayer ground state at $d/l_{\rm B} \simeq 1\mbox{--}2$.
This expectation is in fact correct in the regime of $0< \delta V^{(1)}_1/V^{(1)}_1 \lesssim 0.1$, where the square of overlap is maximal at $d/l_{\rm B} \simeq 2$.
A surprise is that the exact $5/2$ state at the pure Coulomb point is connected with the antisymmetrized bilayer ground state at $d/l_{\rm B} \gg 1$. 

Interestingly, while maintaining high overlap with the exact $5/2$ state at the Coulomb point, the antisymmetrized bilayer ground state at $d/l_{\rm B} \gg 1$ gets separated from the regime of the MR Pfaffian state under moderate anisotropy, which is implemented via non-unity aspect ratio, say, $r_a \simeq 0.8 \mbox{--} 0.9$. 
See Fig.~\ref{fig:Phase_diagram}~(c) and (g), where $r_a=0.8$. 
It is important to note that the non-unity aspect ratio mimics physical anisotropy in finite-size systems, as explained in detail in Sec.~\ref{sec:Anisotropic_instability}.
The above separation suggests that the antisymmetrized bilayer ground state at $d/l_{\rm B} \gg 1$ may belong to a different universality class of phase from the MR Pfaffian state, which is induced via anisotropic instability. 
It is worth mentioning that the MR Pfaffian state itself can be generalized to an anisotropic version via continuous deformation of the $z_i -z_j$ factor~\cite{PhysRevB.85.115308}, which leaves such an anisotropic version in the same universality class as the MR Pfaffian state.

\begin{figure}[]
\includegraphics[width=1\columnwidth]{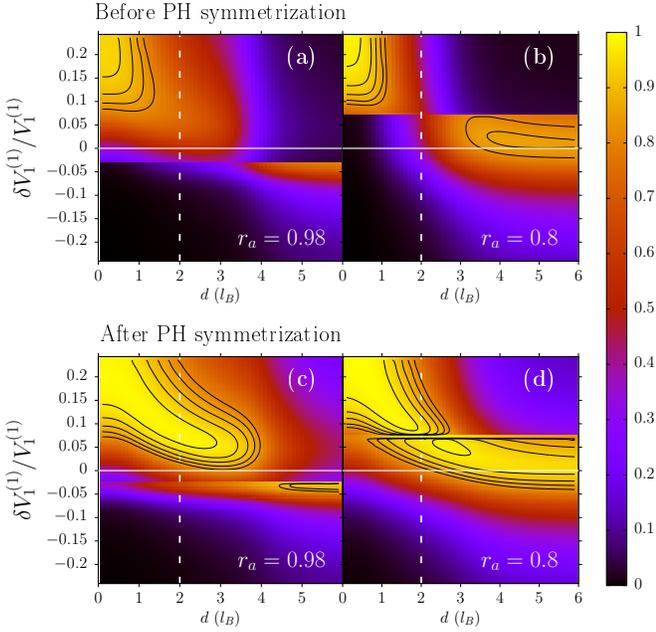}
\caption{
Similar plot to Fig.~\ref{fig:Phase_diagram} for the square of overlap between the exact $5/2$ state, $\Psi_{5/2}[\delta V^{(1)}_1]$, and the antisymmetrized bilayer ground state, ${\cal A}\Psi_{\rm BQH}[d]$,
at ${\bf K}=(0,6)$, whose ground state at $\delta V_1^{(1)}=0$ is globally the third lowest energy state at $r_a=0.98$ and the second lowest energy state at $r_a=0.80$. 
Note that the global ground state occurs at ${\bf K}=(6,6)$.
See Fig.~\ref{fig:Energy_spectrum} for the exact 5/2 energy spectra at $\delta V_1^{(1)}=0$.
As before, several contour lines are drawn, along which the square of overlap takes the constant values of $81\%$, $85\%$, $89\%$, $93\%$, and $97 \%$ from low to high values.
Note that the PH parity of the exact 5/2 state changes at the first-order-like phase boundary.
}
\label{fig:Phase_diagram_excited}
\end{figure}

It turns out that the separation between the MR Pfaffian state and the antisymmetrized bilayer ground state at $d/l_{\rm B} \gg 1$ manifests itself more dramatically at an excited or quasidegenerate pseudomomentum sector, say, ${\bf K}=(0,6)$. 
As mentioned previously, there is a threefold quasidegeneracy among ${\bf K}=(6,6)$, $(6,0)$, and $(0,6)$. 
Around the Coulomb point, the ground state at ${\bf K}=(0,6)$ is globally the third lowest energy state at $r_a=0.98$ and the second lowest energy state at $r_a=0.8$.
Figure~\ref{fig:Phase_diagram_excited} shows the square of overlap between $\Psi_{5/2}[\delta V^{(1)}_1]$ and ${\cal A}\Psi_{\rm BQH}[d]$ at ${\bf K}=(0,6)$, which displays that there is a clear first-order-like phase transition between the exact $5/2$ states occurring at small and large $\delta V^{(1)}_1/V^{(1)}_1$. 
While the exact phase boundary depends on $r_a$, this clear phase transition reveals that the antisymmetrized bilayer ground state at $d/l_{\rm B} \gg 1$ is completely separated from the MR Pfaffian state since they are adiabatically connected to the exact $5/2$ state  occurring at small and large $\delta V^{(1)}_1/V^{(1)}_1$, respectively.

Switching back to Fig.~\ref{fig:Phase_diagram}, we note that the high-overlap trajectory does not go into the regime of negative $\delta V^{(1)}_1/V^{(1)}_1$, where the antisymmetrized bilayer ground state is completely disconnected from the exact ground state in the half-filled SLL.
Intriguingly, this regime is exactly where the exact ground state in the half-filled SLL is predicted to be the stripe state~\cite{PhysRevLett.84.4685, PhysRevB.80.241311}.
In other words, the bilayer mapping does not exist for the stripe state.
This means that, occurring at the pure Coulomb point, the the antisymmetrized bilayer ground state at $d/l_{\rm B} \gg 1$ is neither the isotropic paired state such as the MR Pfaffian state nor the anisotropic compressible state such as the stripe state. 
Incidentally, in contrast to the $N=10$ system~\cite{PhysRevLett.84.4685}, 
the transition between the stripe and the paired regimes appears to be smooth in the $N=12$ system.

Now, let us examine more closely how the square of overlap behaves as a function of aspect ratio, $r_a$.
Varying $r_a$ from unity to other values breaks the discrete rotation symmetry of the unit cell from $C_4$ to $C_2$~\cite{PhysRevLett.109.266806,PhysRevB.86.085129,PhysRevB.86.035122,PhysRevB.87.235128}, which can be used to investigate the possibility of spontaneous symmetry breaking in finite-size system studies~\cite{PhysRevLett.83.1219,PhysRevLett.85.5396}. 
Normally, in the study of the FQHE states, $r_a$ is varied to test if the energy spectrum is robust. 
If so, it would suggest that the ground state is incompressible.   
By contrast, the energy spectrum for the compressible states such as the CF sea and the stripe state would exhibit substantial changes as a function of $r_a$. 
In the context of the half-filled SLL, varying $r_a$ can tune energy splitting of the threefold quasidegenerate ground states~\cite{PhysRevB.80.241311,PhysRevB.78.155308}.


\begin{figure}[t]
\includegraphics[width=1\columnwidth]{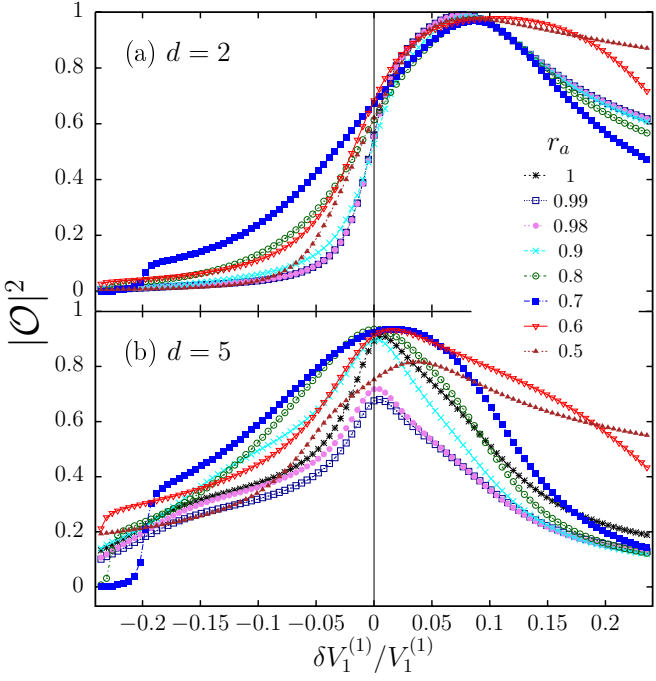}
\caption{
Square of overlap, $|{\cal O}|^2$, between the exact $5/2$ state, $\Psi_{5/2}[\delta V^{(1)}_1]$, and the PH-symmetry-restored antisymmetrized bilayer ground state, ${\cal S}_{\rm PH} \left\{ {\cal A} \Psi_{\rm BQH} [d] \right\}$, at (a) $d/l_{\rm B}=2$ and (b) $5$ as a function of $\delta V^{(1)}_1/V^{(1)}_1$ for various $r_a$. 
All calculations are performed at pseudomomentum ${\bf K}=(6, 6)$.
}
\label{fig:Aspect_ratio}
\end{figure}

Figure~\ref{fig:Phase_diagram} shows that the overall pattern of the square of overlap does not change much as a function of $r_a$ unless $r_a$ gets too small, for example, $r_a=0.5$.
Upon closer examination, however, one may notice that there is a very intriguing difference in the overlap behavior as a function of $r_a$ between small and large $d/l_{\rm B}$.
For concreteness, see Fig.~\ref{fig:Aspect_ratio}, which shows the square of overlap at (a) $d/l_{\rm B}=2$ and (b) $5$ as a function of $\delta V^{(1)}_1/V^{(1)}_1$ for various $r_a$. 
Note that, here, the square of overlap is taken between the exact ground state in the half-filled SLL and the PH-symmetry-restored antisymmetrized bilayer ground state. 
At $d/l_{\rm B}=2$, where the MR Pfaffian state is relevant, the square of overlap shows very little difference regardless of $r_a$ with maximum occurring at $\delta V^{(1)}_1/V^{(1)}_1 \simeq 0.1$. 
By contrast, at $d/l_{\rm B}=5$, the square of overlap drops abruptly as soon as $r_a$ deviates from unity even by less than $1 \%$.  
Specifically, the square of overlap drops from over $90 \%$ to about $60 \%$ by varying $r_a$ from 1.0 to 0.99 at the pure Coulomb point. 
Interestingly, the square of overlap bounces back upon further reducing $r_a$ and even exceeds its value at $r_a=1$ by the time when $r_a$ reaches $0.8$.
Eventually, for sufficiently small $r_a$, say, $r_a=0.5$, it decreases again and becomes far below its value at $r_a=1$.

Such an abrupt drop of the overlap in the vicinity of $r_a=1$ and a subsequent restoration at lower $r_a$ are quite unexpected and puzzling. 
On one hand, the abrupt drop of the overlap indicates that either the exact ground state in the half-filled SLL or the antisymmetrized bilayer ground state undergoes an abrupt change, while the other does not.  
The bilayer ground state at large $d/l_{\rm B}$ before antisymmetrization is the product state of two CF seas at quarter filling, which is susceptible to anisotropic instability since it is not only compressible in each pseudospin sector, but also subject to additional energy degeneracies due to the negligible interlayer interaction. 
This leads us to expect that the antisymmetrized bilayer ground state might also be susceptible to anisotropic instability.
In this situation, if the exact ground state in the half-filled SLL changes smoothly in the vicinity of unity $r_a$, the overlap can drop abruptly. 
Therefore, the abrupt drop of the overlap can be regarded as a sign of spontaneous symmetry breaking in the antisymmetrized bilayer ground state at large $d/l_{\rm B}$ with respect to the rotational symmetry.

On the other hand, the subsequent restoration of the overlap at smaller $r_a$ suggests that the connection between the two states remains intimate and even gets strengthened at a reasonable degree of higher anisotropy. 
Note that the dropped value of the overlap at $r_a=0.99$ is actually comparable to that of the MR Pfaffian state at the pure Coulomb point. 
A puzzle is why its behavior is not as abrupt as that of the antisymmetrized bilayer ground state at large $d/l_{\rm B}$.

One way of resolving this puzzle is to first accept that the exact $5/2$ state at the Coulomb point, $\Psi_{5/2}[\delta V^{(1)}_1=0]$, is fundamentally the same phase as the antisymmetrized bilayer ground state at large $d/l_{\rm B}$, i.e., ${\cal A}\Psi_{^4{\rm CFS}_\uparrow}\otimes \Psi_{^4{\rm CFS}_\downarrow}$, which is apparently susceptible to anisotropic instability.
Strictly speaking, however, there should be no spontaneous symmetry breaking in finite-size systems. 
Therefore, it is not surprising that $\Psi_{5/2}[\delta V^{(1)}_1=0]$ or any other ground states in finite-size systems change smoothly in the presence of small symmetry breaking terms, which include non-unity $r_a$. 
In this sense, actually, more peculiar is the abrupt change of the antisymmetrized bilayer ground state at large $d/l_{\rm B}$. 
This peculiar property is due to the presence of almost perfect degeneracy within the same pseudomomentum sector, which in turn originates from the negligible interlayer interaction at sufficiently large $d/l_{\rm B}$, say, $d/l_{\rm B} \gtrsim 4$. 
We believe that the $N=12$ system is sufficiently large for ${\cal A}\Psi_{^4{\rm CFS}_\uparrow}\otimes \Psi_{^4{\rm CFS}_\downarrow}$ to exhibit a crossover behavior faithfully approximating the thermodynamic limit. 
Unfortunately, the situation is different for the exact $5/2$ state, for which the $N=12$ system might be below the critical system size.

\begin{figure}[t]
\includegraphics[width=1\columnwidth]{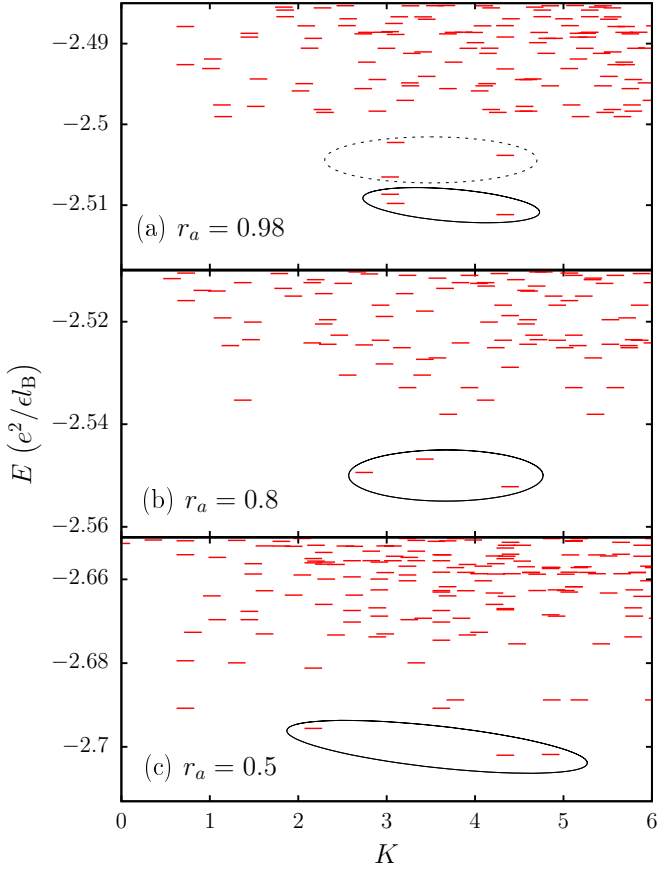}
\caption{
Exact energy spectra of the $N=12$ system in the half-filled SLL at the Coulomb point, i.e., $\delta V^{(1)}_1=0$
for $r_a=$ (a) 0.98, (b) 0.8, and (c) 0.5 as a function of pseudomomentum magnitude $K= [ (2\pi K_x/a)^2+ (2\pi K_y/b)^2 ]^{1/2}$ in units of $1/l_{\rm B}$. 
Note that there is a threefold quasidegeneracy among the ground states in ${\bf K}= (6,6)$, $(6,0)$, and $(0,6)$, which are enclosed by solid ellipses for guide to eye.   
There are additional quasidegenerate excited copies of the ground states at near-unity $r_a$, which are enclosed by a dashed ellipse in (a). 
} 
\label{fig:Energy_spectrum}
\end{figure}

At this point, one may wonder if the energy gap remains stable as $r_a$ deviates from unity. 
To check this, we have computed the exact energy spectra of the $N=12$ system in the half-filled SLL at the pure Coulomb interaction, which are shown in Fig.~\ref{fig:Energy_spectrum} for three difference values of $r_a=$ (a) 0.98, (b) 0.8, and (c) 0.5.
As one can see, the overall structure of energy spectra remains almost invariant up to $r_a=0.8$ with a clear separation between the ground states, which form a threefold quasidegenerate subspace, and the ``continuum'' of excited states.  
This structure is destroyed if $r_a$ becomes too small, as shown in Fig.~\ref{fig:Energy_spectrum}~(c).
Remember that the overall pattern of the square of overlap is also changed substantially at this regime of small $r_a$.

It is interesting to note that, at near-unity $r_a$, there are additional quasidegenerate excited copies of the ground states, which are enclosed by a dashed ellipse in Fig.~\ref{fig:Energy_spectrum}~(a).
A conventional interpretation of these additional quasidegenerate copies is that they are due to the mixing between the MR Pfaffian state and its PH conjugate, or the anti-Pfaffian state~\cite{PhysRevLett.99.236806,PhysRevLett.101.016807,PhysRevB.80.241311,PhysRevB.78.155308}. 
As mentioned previously, there is an exact PH symmetry in the half-filled SLL for the Coulomb interaction, and therefore the MR Pfaffian and anti-Pfaffian states would be degenerate if there were no mixing between the two. 
Apparently, however, there is mixing, which splits the otherwise degenerate doublet into the symmetric and antisymmetric linear combination.
In view of anisotropic instability discussed in the above, we have a different interpretation than this. 
Our interpretation is that the additional quasidegenerate copies could be due to the symmetric and antisymmetric linear combination between the rotational-symmetry broken states, not between the MR Pfaffian and anti-Pfaffian states. 
It is worth mentioning that the threefold quasidegeneracy in the excited states is much more fragile than that in the ground states in the sense that it is highly dependent on the system size; it is not observed at all in the $N=14$ system and partially observed in the $N=16$ system~\cite{PhysRevB.78.155308}.

So far, we have proposed a conjecture that the abrupt drop of the overlap is induced by the spontaneous symmetry breaking of the antisymmetrized bilayer ground state at large $d/l_{\rm B}$ in the presence of a small discrete rotational symmetry breaking term such as non-unity $r_a$.
In the following section, we prove that this conjecture is actually true. 
Specifically, we perform a systematic study on the behavior of the square of overlap as a function of aspect ratio as well as physical anisotropy parameters such as the anisotropy ratio of the band mass~\cite{PhysRevB.85.165318} and the dielectric tensor of the Coulomb interaction~\cite{PhysRevB.86.035122}.

\section{Anisotropic instability}
\label{sec:Anisotropic_instability}

While serving as a useful symmetry-breaking parameter for the discrete rotational symmetry in the torus geometry, the aspect ratio of the unit cell is not a physically meaningful parameter truly valid in the thermodynamic limit.  
Two of the most natural, physically meaningful parameters are the anisotropy ratio of the band mass~\cite{PhysRevB.85.165318} and that of the dielectric tensor of the Coulomb interaction~\cite{PhysRevB.86.035122}.

To see how the band mass and the dielectric tensor anisotropy are implemented in ED, let us begin by writing the Coulomb matrix element for the bilayer QH system, which is similar to Eq.~\eqref{eq:Matrix_element}:
\begin{align}
M^{(n)}_{j_1 j_2 j_3 j_4}
&=\delta'_{j_1\!-j_4,t}\delta'_{j_1+j_2,j_3+j_4}
\nonumber \\
&\times \sideset{}{'}\sum_{\bf q} {V^{\sigma_{j_1},\sigma_{j_2}}_{\bf q}\over 2N_{\Phi}} e^{iq_x(X_{j_1}-X_{j_3})}
 e^{-{q^2\over 2}} L_n^2\left({q^2\over 2}\right),
 \label{eq:BQH_Matrix_element}
\end{align}
where ${\bf q}=(q_x,q_y)=(\frac{2\pi}{a}s, \frac{2\pi}{b} t)$ $\left[s, t\in {\mathbb Z}\right]$ and
$V^{\sigma_{j_1},\sigma_{j_2}}_{\bf q}$ is the pseudospin-dependent interaction potential between electrons in the Landau-gauge orbital $j_1$ and $j_2$.

The band mass anisotropy is implemented via replacing $q$ inside $e^{-{q^2/ 2}} L_n^2\left(q^2/2\right)$ by $(q_x^2/r_m +q_y^2 r_m)^{1/2}$, where $r_m$ is the anisotropy ratio of band mass~\cite{PhysRevB.85.165318}.
Meanwhile, the dielectric tensor anisotropy is implemented via replacing $q$ inside $V^{\sigma_{j_1},\sigma_{j_2}}_{\bf q}$ by $(q_x^2/r_d +q_y^2 r_d)^{1/2}$, where $r_d$ is the anisotropy ratio of dielectric tensor~\cite{PhysRevB.86.035122}.
For simplicity, we study anisotropy effects by varying $r_m$ and $r_d$ in the square unit cell with $a=b$.
It is interesting to note that varying the aspect ratio is formally equivalent to replacing $q$ in both  $e^{-{q^2/ 2}} L_n^2\left(q^2/2\right)$ and $V^{\sigma_{j_1},\sigma_{j_2}}_{\bf q}$ by $(q_x^2/r_a +q_y^2 r_a)^{1/2}$.
Therefore, the Hamiltonian at aspect ratio, $r_a$, can be mapped onto that of the anisotropic model with both the band mass and the dielectric tensor anisotropy ratio being equal to $r_a$ in the square unit cell.

\begin{figure}[]
\includegraphics[width=1\columnwidth]{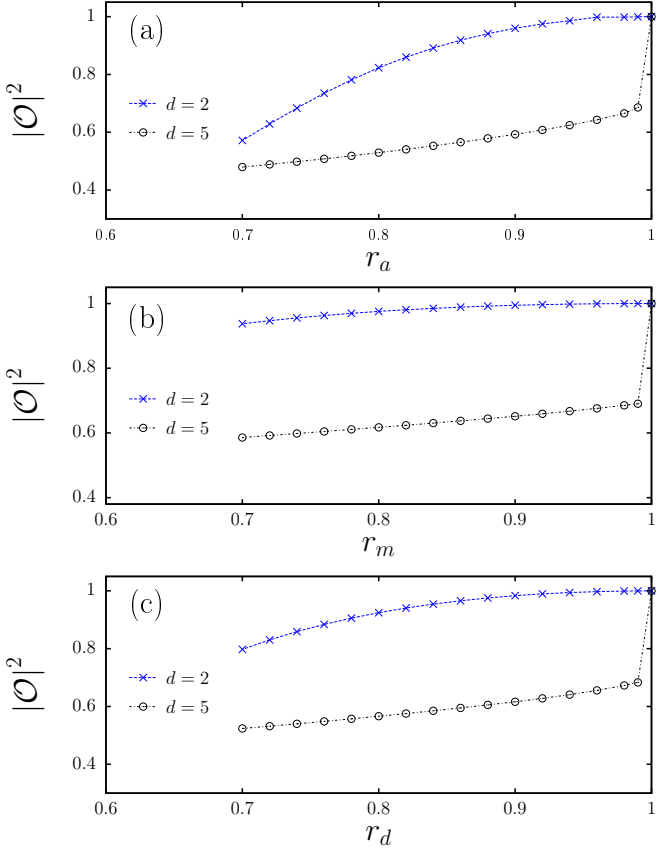}
\caption{ 
Square of ``self-overlap,'' $|{\cal O}|^2$, between the PH-symmetry-restored antisymmetrized bilayer ground states with respect to the variation of (a) the aspect ratio of the unit cell, $r_a$, (b) the anisotropy ratio of band mass, $r_m$, and (c) the anisotropy ratio of dielectric tensor, $r_d$, at $d/l_{\rm B}=$ 2 and 5.
Specifically, (a) shows the square of overlap, $|{\cal O}|^2$, between the PH-symmetry-restored antisymmetrized bilayer ground state in the square unit cell, ${\cal S}_{\rm PH}\{{\cal A}\Psi_{\rm BQH}[d;r_a=1]\}$, and that in the rectangular unit cell at general $r_a$, ${\cal S}_{\rm PH}\{{\cal A}\Psi_{\rm BQH}[d;r_a]\}$. 
(b) and (c) show the similar square of overlap as a function of $r_m$ and $r_d$, respectively.  
Note that, in the study of the band mass as well as the dielectric tensor anisotropy, all states are obtained in the square unit cell.
}
\label{fig:Anisotropy}
\end{figure}

While the rotational invariance is preserved in the thermodynamic limit when the two anisotropic effects are congruent~\cite{2009arXiv0906.1854H,PhysRevB.84.085316}, they are not necessarily compensated exactly in finite-size systems. 
As a consequence, if the ground state responds differently to each variation of $r_m$ and $r_d$, changing the aspect ratio can give rise to results qualitatively similar to those obtained from the variation of $r_m$ and/or $r_d$.  
This is the reason why the aspect ratio dependence of the overlap, which is studied in Sec.~\ref{sec:Bilayer_mapping}, can provide an important clue to the effects of anisotropy.
We think that this relation between the real anisotropy and the aspect ratio underlies the occurrence of similar solid phases in the large anisotropy~\cite{PhysRevB.86.035122} as well as the thin torus limit~\cite{PhysRevB.50.17199,PhysRevLett.92.096401,PhysRevLett.95.266405,PhysRevLett.94.026802} at $\nu=1/3$.

With this background, let us examine the square of ``self-overlap'' between the PH-symmetry-restored antisymmetrized bilayer ground states with respect to the variation of aspect ratio, $r_a$, the band mass anisotropy ratio, $r_m$, and the dielectric tensor anisotropy ratio, $r_d$, which are shown in Fig.~\ref{fig:Anisotropy}~(a), (b), and (c), respectively.
Specifically, in Fig.~\ref{fig:Anisotropy}~(a), the square of overlap, $|{\cal O}|^2$, is taken between the PH-symmetry-restored antisymmetrized bilayer ground state in the square unit cell, ${\cal S}_{\rm PH}\{{\cal A}\Psi_{\rm BQH}[d;r_a=1]\}$, and those in the rectangular unit cell with general $r_a$, ${\cal S}_{\rm PH}\{{\cal A}\Psi_{\rm BQH}[d;r_a]\}$. 
That is to say, we compute 
\begin{align}
| \langle {\cal S}_{\rm PH}\{{\cal A}\Psi_{\rm BQH}[d;r_a=1]\} | {\cal S}_{\rm PH}\{{\cal A}\Psi_{\rm BQH}[d;r_a]\} \rangle |^2.
\end{align}
Fig.~\ref{fig:Anisotropy}~(b) and (c) shows the similarly defined square of overlap
as a function of $r_m$ and $r_d$, respectively.

It is worth mentioning that a straight computation of the overlap between the ground states obtained at different $r_a/r_m$ suffers from a conceptual issue that the basis states themselves are deformed as a function of $r_a/r_m$.  
Here, we ignore this issue by assuming that the two basis states are the same if they have the same sequence of orbital quantum numbers.
Under this assumption, we can concentrate on the many-body response of the ground state on anisotropy, which is imbedded in the amplitudes of the basis states.

One of the first things to notice is that there are very little differences among the dependencies of the overlap on all three anisotropy ratios.   
At $d/l_{\rm B}=2$, the square of overlap changes smoothly.
Meanwhile, at $d/l_{\rm B}=5$, the square of overlap drops abruptly as soon as the anisotropy ratios deviate from unity. 
This result is consistent with our conjecture that the antisymmetrized bilayer ground state at large $d/l_{\rm B}$ is susceptible to anisotropy instability, which manifests itself as a sudden change of the wave function in the presence of small symmetry breaking terms.

To verify more explicitly that the abrupt drop of the overlap near $r_a=1$ is indeed a consequence of spontaneous (discrete) rotational symmetry breaking, we construct the following trial wave function:
\begin{align}
\Psi_{\pm}[d;r_a]  = \frac{1}{\cal N} {\cal S}_{\rm PH} \{
{\cal A} \Psi_{\rm BQH}[d;r_a] \pm {\cal A} \Psi_{\rm BQH}[d,r_a^{-1}]
\} ,
\end{align}
where $1/{\cal N}$ is the normalization constant.
The phase of each component is fixed by using the following convention:
\begin{align}
&\langle {\cal S}_{\rm PH} \{ {\cal A} \Psi_{\rm BQH}[d;r_a=1]\} | {\cal S}_{\rm PH} \{ {\cal A} \Psi_{\rm BQH}[d;r_a]\} \rangle     
\nonumber \\
= & \langle {\cal S}_{\rm PH} \{ {\cal A} \Psi_{\rm BQH}[d;r_a=1]\} |    {\cal S}_{\rm PH} \{ {\cal A} \Psi_{\rm BQH}[d,r_a^{-1}] \} \rangle ,
\end{align}
which fixes only the phase since both ${\cal S}_{\rm PH}\{{\cal A} \Psi_{\rm BQH}[d;r_a]\}$ and ${\cal S}_{\rm PH}\{{\cal A} \Psi_{\rm BQH}[d;r_a^{-1}]\}$ have exactly the same absolute value of overlap with ${\cal S}_{\rm PH}\{{\cal A} \Psi_{\rm BQH}[d;r_a=1]\}$.
The rationale behind the above trial wave function is that ${\cal S}_{\rm PH} \{ {\cal A} \Psi_{\rm BQH}[d;r_a^{-1}] \}$ can be regarded as a conjugate of ${\cal S}_{\rm PH} \{ {\cal A} \Psi_{\rm BQH}[d;r_a] \}$ with respect to discrete rotation by $90^\circ$ that exchanges the role of the $x$ and $y$ directions. 
Then, by making a symmetric or antisymmetric combination of the two states, one can construct a ``rotational-symmetry-restored'' trial state at given $r_a$.

\begin{figure}[]
\includegraphics[width=1\columnwidth]{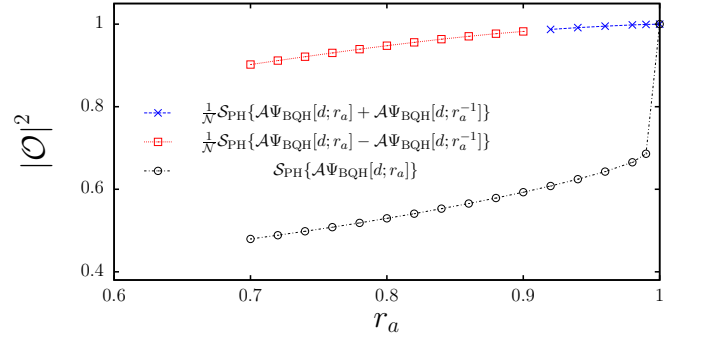}
\caption{ 
Square of overlap, $|{\cal O}|^2$, between the PH-symmetry-restored antisymmetrized bilayer ground state at unity $r_a$, ${\cal S}_{\rm PH}\{{\cal A}\Psi_{\rm BQH}[d;r_a=1]\}$, and its ``rotational-symmetry-restored'' version at general $r_a$, $\frac{1}{\cal N}  {\cal S}_{\rm PH} \{ {\cal A}\Psi_{\rm BQH}[d;r_a] \pm {\cal A}\Psi_{\rm BQH}[d;r_a^{-1}]\}$, where $1/{\cal N}$ is the normalization constant.
Here, $d/l_{\rm B}=5$.
This provides a compelling piece of evidence for spontaneous breaking of the discrete rotational symmetry at large $d/l_{\rm B}$. 
}
\label{fig:SSB}
\end{figure}

In our current situation, spontaneous breaking of the discrete rotational symmetry means that the exact ground state at unity $r_a$, ${\cal S}_{\rm PH}\{{\cal A}\Psi_{\rm BQH}[d;r_a=1]\}$, is composed of both ${\cal S}_{\rm PH}\{{\cal A} \Psi_{\rm BQH}[d;r_a]\}$ and ${\cal S}_{\rm PH}\{{\cal A} \Psi_{\rm BQH}[d;r_a^{-1}]\}$ at $r_a$ very close, not exactly equal to unity, which are rotational-symmetry-broken partners to each other.
Note that ${\cal S}_{\rm PH}\{{\cal A} \Psi_{\rm BQH}[d;r_a]\}$ and ${\cal S}_{\rm PH}\{{\cal A} \Psi_{\rm BQH}[d;r_a^{-1}]\}$ are not necessarily orthogonal, but certainly linearly independent. 
Under this assumption, we can now explain the abrupt change in the square of overlap, which is induced by the abrupt collapse of the ground state wave function from ${\cal S}_{\rm PH}\{{\cal A}\Psi_{\rm BQH}[d;r_a=1]\}$ to one of its components, say, ${\cal S}_{\rm PH}\{{\cal A} \Psi_{\rm BQH}[d;r_a]\}$. 
If this explanation is correct, one can then make a subsequent prediction that $\Psi_{\pm}[d;r_a]$ evolves smoothly as a function of $r_a$ since it contains both ${\cal S}_{\rm PH}\{{\cal A} \Psi_{\rm BQH}[d;r_a]\}$ and ${\cal S}_{\rm PH}\{{\cal A} \Psi_{\rm BQH}[d;r_a^{-1}]\}$.

Figure~\ref{fig:SSB} shows that this prediction is precisely fulfilled. 
As predicted, at $d/l_{\rm B}=5$, the square of overlap between ${\cal S}_{\rm PH}\{{\cal A}\Psi_{\rm BQH}[d;r_a=1]\}$ and $\Psi_{+}[d;r_a]$ shows a smooth behavior near $r_a=1$.
Interestingly, the square of overlap experiences a phase transition at $r_a \simeq 0.9$, where the role of $\Psi_{+}[d;r_a]$ is replaced by $\Psi_{-}[d;r_a]$.
In other words, the square of overlap between ${\cal S}_{\rm PH}\{{\cal A}\Psi_{\rm BQH}[d;r_a=1]\}$ and $\Psi_{-}[d;r_a]$ continues to follow the smooth curve extending that between ${\cal S}_{\rm PH}\{{\cal A}\Psi_{\rm BQH}[d;r_a=1]\}$ and $\Psi_{+}[d;r_a]$.
While we do not know the origin of this phase transition, the main point is that the ground state undergoes an abrupt collapse of the wave function near $r_a=1$, which is consistent with spontaneous breaking of the discrete rotational symmetry.

\section{Conclusion}
\label{sec:Conclusion}

In this work, it is shown that the exact $5/2$ state at a given Haldane pseudopotential variation, $\delta V^{(1)}_1$, is intimately connected with the antisymmetrized bilayer ground state at a corresponding interlayer distance, $d$, via one-to-one mapping, which we call the bilayer mapping.
One of the most important discoveries in this work is that the bilayer mapping connects the exact $5/2$ state occurring at the Coulomb interaction with the antisymmetrized bilayer ground state at $d/l_{\rm B} \gg 1$, not at $d/l_{\rm B} \simeq 1\mbox{--}2$, where the MR Pfaffian state is relevant.

Microscopically, the antisymmetrized bilayer ground state at $d/l_{\rm B} \gg 1$ is described by the antisymmetrized product state of two CF seas at quarter filling with pseudospin up and down, ${\cal A}\Psi_{^4{\rm CFS}_\uparrow}\otimes \Psi_{^4{\rm CFS}_\downarrow}$.
Since the product state of two CF seas at quarter filling is not only compressible in each pseudospin sector, but also subject to additional energy degeneracies due to the negligible interlayer interaction, it is natural to expect that its antisymmetrized version might be susceptible to anisotropic instability.
In accordance with this expectation, the antisymmetrized bilayer ground state at $d/l_{\rm B} \gg 1$ exhibits an abrupt collapse of the wave function from the isotropic to the anisotropic state, as soon as the aspect ratio, $r_a$, deviates from unity. 
Considering with the fact that (i) the exact $5/2$ state is gapped at the Coulomb interaction and (ii) the the antisymmetrized bilayer ground state at $d/l_{\rm B} \gg 1$ is susceptible to anisotropic instability, 
we conclude that the exact $5/2$ state might be better described by the anisotropic paired QH state with $p_x$ or $p_y$-wave symmetry than by the MR Pfaffian/anti-Pfaffian states with $p_x \pm i p_y$-wave symmetry.

A remaining question is how the antisymmetrization operation can make the product state of two CF seas at quarter filling gapped.
While a precise understanding is not yet obtained, we would like to suggest a plausible explanation.   
As mentioned in Sec.~\ref{sec:ABGS}, the ``pairing strength'' is related with the discrepancy between the intralayer and interlayer interaction between electrons,
in which sense pairing is the strongest at $d/l_{\rm B} \gg 1$. 
While this may sound strange at first, remember that pairing is formed not between electrons, but between composite fermions carrying even numbers of vortices, which is nothing but quantized correlation holes.
Since there is no repulsive correlation between different layers at $d/l_{\rm B} \gg 1$, composite fermions do not carry any correlation holes with respect to electrons in the opposite layer. 
In this sense, composite fermions can be thought as having a relative ``attractive'' interaction between different layers. 
A problem is that the simple product state of two CF seas at quarter filling is gapless since composite fermions are not actually paired up to form a Cooper pair.   
In other words, composite fermions could form a Cooper pair with any one of them, but did not yet choose the partner.
We conjecture that antisymmetrization plays a role of forcing them to choose the partner.

Finally, it is interesting to note that there is a functional similarity between the antisymmetrization operation and the Gutzwiller projection, both of which remove doubly occupied configurations from a given state.
In this context, it is worthwhile to mention that the Gutzwiller-projected product state of two Fermi seas with spin up and down had been studied in the context of the one-dimensional $S=1/2$ Heisenberg model. 
Actually, the nearest-neighbor one-dimensional $S=1/2$ Heisenberg model is exactly solved with the ground state given by the Bethe ansatz solution.  
Interestingly, the exact Bethe ansatz solution is very well approximated by the Gutzwiller-projected product state of two Fermi seas with spin up and down, which, in addition to the closeness in energy, shows a power-law spin-spin correlation function similar to the exact result. 
Furthermore, Haldane~\cite{PhysRevLett.60.635} and Shastry~\cite{PhysRevLett.60.639} showed that the Gutzwiller-projected product state of two Fermi seas was the exact ground state of the one-dimensional $S=1/2$ Heisenberg model with $1/r^2$ exchange coupling, which can be also identified as Anderson's resonating valence bond (RVB) state~\cite{RVB}. 
It is inspiring to think that there is an underlying connection between the paired QH state and the RVB state.

\acknowledgements
%
%
The authors gratefully acknowledge critical comments from Jainendra K. Jain and Steven H. Simon.
We are also thankful to Hantao Lu, Seung Ki Baek, Kenji Hashimoto, Jaeyoon Cho, Beom Hyun Kim, Suk Bum Chung, Jun-Won Rhim, and Woo-Ram Lee for useful discussions. 
JSJ appreciates the hospitality of Yong Baek Kim at University of Toronto, where parts of this work were performed. 
We thank KIAS center for Advanced Computation for providing computing resources. 
This work was supported by the Supercomputing Center/Korea Institute of Science and Technology Information with supercomputing resources including technical support (KSC-2013-C3-050 and KSC-2014-C3-033).

\appendix

\section{Particle-hole symmetrization}
\label{appendix:PH_symmetrization}

The particle-hole (PH) symmetrization is implemented in the torus geometry by adding or subtracting each basis state with its PH transformed state, which is obtained via the combination of two operations with one being a straight PH transformation and the other being an antiunitary reflection~\cite{PhysRevLett.84.4685}. 
The straight PH transformation is performed via PH reversal of the occupation numbers, followed by rotation by $180^\circ$ and complex conjugation of the amplitudes. 
As pointed by Rezayi and Haldane~\cite{PhysRevLett.84.4685}, a problem is that the straight PH transformation alone is an antiunitary operation, which does not support symmetry-related eigenstates. 
To make the symmetry unitary, it is necessary to combine the straight PH transformation with an antiunitary reflection, which is in turn composed of reflection and complex conjugation of the amplitudes.

It is important to note that, when applied alone, PH reversal of the occupation numbers changes the total momentum of each basis state. 
Basis states are usually denoted as binary sequences composed of 0 (empty state) and 1 (occupied state) for each orbital of the Landau gauge momentum, $k$, whose value ranges between 1 and $N_{\Phi}$ in units of $2\pi/b$.  
In this situation, the total momentum of a given basis state can be written as $K_{\rm tot}=\sum_{i=1}^N k_i$, where $k_i$ denotes the Landau gauge momentum of the $i$-th occupied orbital.
Then, the total momentum of the PH-reversed basis state is given by $K^{\rm PH}_{\rm tot} = N_\Phi(N_\Phi+1)/2 -K_{\rm tot}$, which is not necessarily equal to $K_{\rm tot}$. 
To overcome this problem, one would like to devise a PH transformation that leaves the total momentum of each basis state {\it invariant}.
Such a PH transformation can be obtained by combining PH reversal with rotation by $180^\circ$, which is implemented by the following transformation: $k_i \rightarrow N_\Phi +1 -k_i$.  
That is to say, rotation by $180^\circ$ gives rise to $K^{\rm PH}_{\rm tot} \rightarrow N_\Phi(N_\Phi+1)/2 -\sum_{i=1}^N (N_\Phi +1 -k_i)= N_\Phi(N_\Phi+1)/2 -N(N_\Phi+1) +\sum_{i=1}^N k_i = K_{\rm tot}$, where $N=N_\Phi/2$ at half filling.
Therefore, the total momentum is restored.

Finally, note that the role of the antiunitary reflection operation is twofold. 
The first role is to cancel the complex conjugation operation imposed as a part of the straight PH transformation.
The second is to set restrictions on the allowed values for the total momentum.
Specifically, the total momentum, ${\bf K}$, itself should be invariant with respect to reflection so that it can return to itself after reflection.  
In the torus geometry, those values are ${\bf K}= (N/2,N/2)$, $(N/2,0)$, $(0,N/2)$, and $(0,0)$.
In this work, we focus on ${\bf K}= (N/2,N/2)$ since the ground state in the half-filled second Landau level 
almost always occurs in this momentum sector at $N=12$ except for few sporadic parameter regimes.

\bibliography{Reference}

\end{document}